\documentclass[english,12pt, draftclsnofoot, onecolumn]{IEEEtran}
\usepackage[T1]{fontenc}
\usepackage[latin9]{inputenc}
\usepackage{color}
\usepackage{verbatim}
\usepackage{amsmath}
\usepackage{amsthm}
\usepackage{amssymb}
\usepackage{graphicx}
\usepackage{esint}

\makeatletter

\providecommand{\tabularnewline}{\\}

  \theoremstyle{plain}
  \newtheorem{prop}{\protect\propositionname}
  \theoremstyle{plain}
  \newtheorem{cor}{\protect\corollaryname}
  \theoremstyle{plain}
  \newtheorem{lem}{\protect\lemmaname}
  \theoremstyle{plain}
  \newtheorem{thm}{\protect\theoremname}

\usepackage{cite}
\usepackage{tensor}

\@ifundefined{showcaptionsetup}{}{%
 \PassOptionsToPackage{caption=false}{subfig}}
\usepackage{subfig}
\makeatother

\usepackage{babel}
\providecommand{\corollaryname}{Corollary}
\providecommand{\lemmaname}{Lemma}
\providecommand{\propositionname}{Proposition}
\providecommand{\theoremname}{Theorem}

\begin{document}

\title{Information-Centric Offloading in Cellular Networks with Coordinated
Device-to-Device Communication}

\author{\medskip{}
}

\author{{\normalsize{}Asma Afzal, Syed Ali Raza Zaidi, Des McLernon and Mounir
Ghogho}{\small{}}\thanks{{\small{}The authors are with the School of Electronic and Electrical
Engineering, University of Leeds, United Kingdom.}\protect \\
{\small{}M. Ghogho is also affliated with the University of Rabat,
Morocco.}\protect \\
{\small{}Email: \{elaaf, s.a.zaidi, d.c.mclernon, m.ghogho\}@leeds.ac.uk}}}
\maketitle
\begin{abstract}
\textcolor{black}{In this paper, we develop a comprehensive analytical
framework for cache enabled cellular networks overlaid with coordinated
device-to-device (D2D) communication. We follow an approach similar
to LTE Direct, where the base station (BS) is responsible for establishing
D2D links. We consider that an arbitrary requesting user is offloaded
to D2D mode to communicate with one of its }\textit{\textcolor{black}{`k'}}\textcolor{black}{{}
closest D2D helpers within the macrocell subject to content availability
and helper selection scheme. We consider two different D2D helper
selection schemes: 1) uniform selection (US), where the D2D helper
is selected uniformly and 2) nearest selection (NS), where the nearest
helper possessing the content is selected. Employing tools from stochastic
geometry, we model the locations of BSs and D2D helpers using independent
homogeneous Poisson point processes (HPPPs). We characterize the D2D
mode probability of an arbitrary user for both the NS and US schemes.
The distribution of the distance between an arbitrary user and its
}\textit{\textcolor{black}{i}}\textcolor{black}{th neighboring D2D
helper within the macrocell is derived using disk approximation for
the Voronoi cell, which is shown to be reasonably accurate. We fully
characterize the overall coverage probability and the average ergodic
rate of an arbitrary user requesting a particular content. We show
that significant performance gains can be achieved compared to conventional
cellular communication under both the NS and US schemes when popular
contents are requested and NS scheme always outperforms the US scheme.
Our analysis reveals an interesting trade off between the performance
metrics and the number of candidate D2D helpers }\textit{\textcolor{black}{`k'}}\textcolor{black}{.
We conclude that enhancing D2D opportunities for the users does not
always result in better performance and the network parameters have
to be carefully tuned to harness maximum gains. }
\end{abstract}

\section{Introduction}

\IEEEPARstart{U}{biquitous} devices such as smart phones and tablets
have fueled the demand for data intensive applications, including
ultra high-definition video streaming, social networking and e-gaming.
This puts significant pressure on the traditional cellular networks,
which are not designed to support such high data rates and reliability
requirements. As a consequence, current research on fifth generation
(5G) wireless networks is geared towards developing intelligent ways
of data dissemination by deviating from the traditional host centric
network architecture to a more versatile information centric architecture. 

Caching in the IP networks has already proved to be a promising way
to reduce the overhead of backhaul communication. Borrowing from these
principles, caching at the edge of the cellular networks potentially
reduces the backhaul access cost in terms of capacity, latency and
energy consumption by turning memory into bandwidth \cite{bastug2014living}.
Recent observations have indicated that the data traffic consists
of a lot of duplications of multimedia content requested by various
users in the same vicinity\cite{woo2013comparison}. Therefore, users
can leverage this trend to their advantage by accessing information
pre-downloaded by their neighboring users using Direct device-to-device
(D2D) communication. D2D communication is a promising technique to
improve the coverage and throughput of cellular networks \cite{lin2014overview}.
Mobile users in close physical proximity can exchange popular files
without the intervention of the base station (BS). This not only offloads
the burden of duplicate transmissions from the BS, but it also provides
higher rates due to short range D2D communication \cite{malandrino2014toward}.
Several techniques have been proposed to materialize the concept of
integrating D2D communication with cellular networks. The major design
questions are: Should D2D communication operate in the cellular uplink
or downlink, licensed spectrum or unlicensed spectrum, and in the
licensed spectrum should it be underlay or overlay, coordinated by
the BS or uncoordinated? In this paper, we focus on coordinated in-band
overlay D2D communication in the cellular downlink, where a macro
base station (MBS) establishes, manages and arbitrates a D2D connection
\cite{fodor2012design}. The reader is referred to a detailed discussion
of the other D2D implementation techniques in \cite{asadi2014survey}
and the references therein. 

So, the novelty of this paper is as follows. We propose a new information-centric
offloading mechanism, whereby the MBS maintains a record of the previously
downloaded contents by the users in its long term-coverage region.
Based on this information, the MBS schedules a D2D link between a
user and one of its $k$ neighboring D2D helpers subject to the content
availability and helper selection scheme. These D2D helper devices
can be considered as users which are not receiving any data form the
MBS in the current radio frame and can transmit their data. We consider
two different helper selection schemes, namely, 1) nearest helper
selection (NS): where the MBS selects the D2D helper closest to the
user possessing the requested content and 2) uniform selection (US):
where the MBS uniformly selects a D2D helper first and checks for
content availability later. The estimation of a user's location is
much accurate thanks to the in-built GPS and location apps in smart
phones. Fig. \ref{fig:SystemModel} displays a simple example of the
scenario under consideration. The MBS examines its records for the
arrived content requests and schedules possible D2D transmissions.
Here, User\#1 is served by its second nearest D2D helper, while User\#2
is served by the MBS as none of its $k$ neighboring helpers have
the content. 

\textcolor{black}{We analyze such a system with the help of }stochastic
geometry to quantify the performance improvement compared to conventional
and cache enabled single tier cellular networks. Stochastic geometry
has recently emerged as a powerful tool to accurately analyze the
performance of large scale cellular networks \cite{haenggi2012stochastic}.
We make use of the Poisson point process (PPP) assumption in modeling
the locations of the MBSs and D2D to derive tractable expressions
for various performance metrics. 

The main contributions of this article are summarized as follows.
\begin{itemize}
\item For the information-centric offloading paradigm, we consider that
both the D2D helpers and BSs are equipped with caches and an arbitrary
user requests a certain content based on its popularity. It is important
to note that in this work, we focus on devising efficient data dissemination
techniques for a given content placement strategy. Based on the content
placement strategy, helper selection schemes and other caching parameters,
we derive expressions for the probability that an arbitrary user is
served in D2D mode for both the NS and US schemes. We obtain bounds
on this probability and study its behavior as the number of candidate
helpers $k$ grows. 
\item With the help of our stochastic geometry framework, we derive the
distribution of distance between an arbitrary user and its $i$th
nearest D2D helper within the cell using a disk approximation for
a Voronoi cell. We show that this approximation is fairly accurate
for various values of $i$ and compare it with the distribution of
distance between the requesting user and its $i$th nearest D2D helper
not necessarily present inside the cell. We investigate the conditions
in which our derived distribution reduces to the unconstrained case.
\item We characterize the coverage probability for individual D2D links
and the probability that an arbitrary user is in coverage when it
requests a particular content and operates in D2D mode with the NS
and US schemes. We also validate our results with network simulations.
\item We explore the two important performance metrics, including the overall
coverage probability and the average rate experienced by an arbitrary
user requesting a particular content $c$. We show that there exists
an optimal number of candidate D2D helpers $k$ which maximize the
overall coverage and the average rate. The optimal $k$ maximizing
the coverage probability is independent of caching parameters or the
requested content and only depends on the network parameters. However,
that is not the case for the optimal $k$ maximizing the average rate.
We also show that high performance gains can be harnessed compared
to conventional cellular communication when the most popular contents
are requested. 
\begin{figure}
\begin{centering}
\includegraphics[scale=0.7]{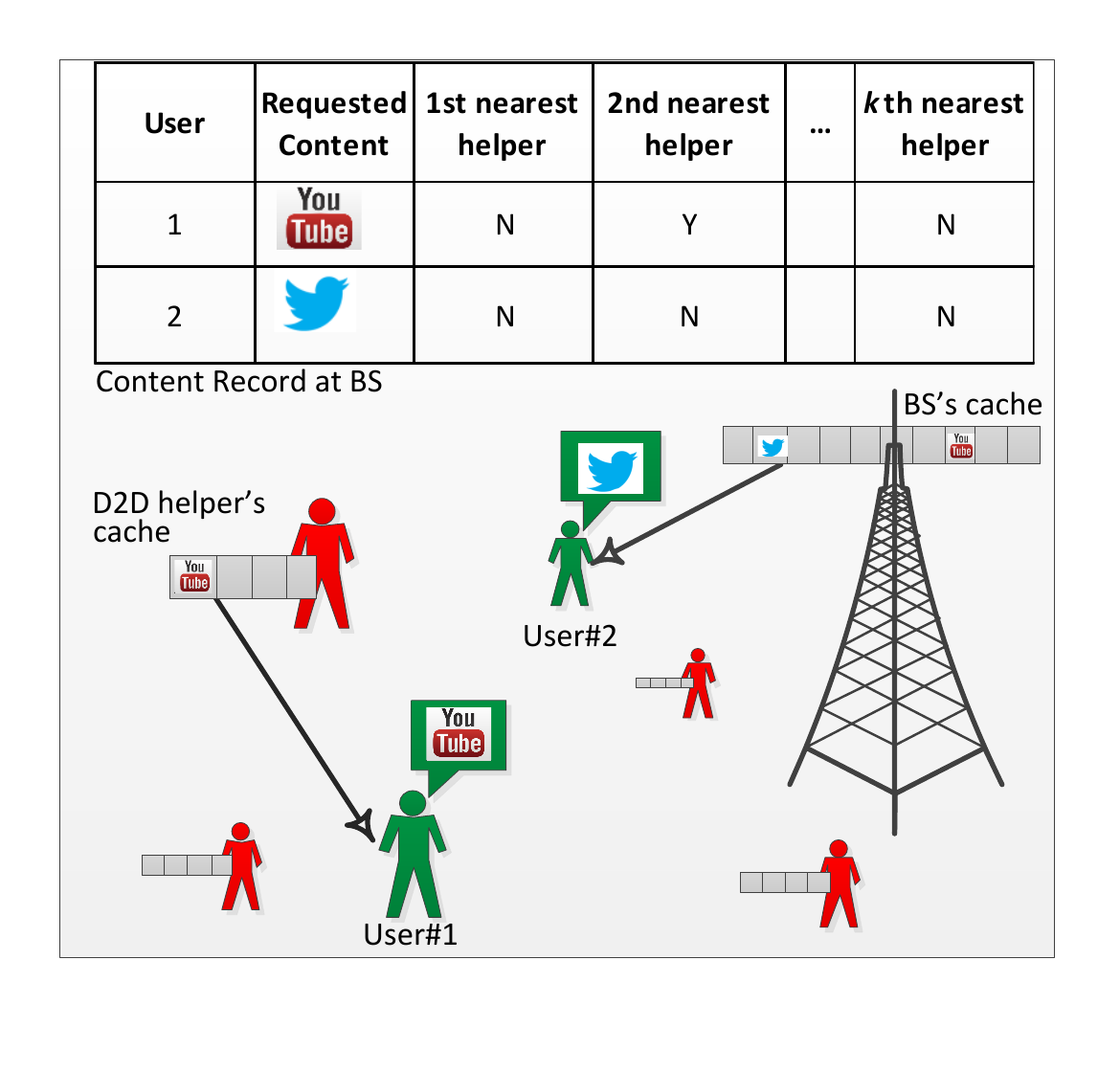}
\par\end{centering}
\centering{}\caption{Illustration of cache-enabled coordinated D2D network. The MBS pairs
the requesting users with one of their $k$ neighbors depending on
the content availability and helper selection scheme. If none of the
$k$ neighbors have the content, the MBS serves the user itself. \label{fig:SystemModel}}
\end{figure}
\end{itemize}

\subsection{Related Work}

Characterization of the performance of caching enabled cellular networks
has been widely studied in \cite{ji2014fundamental,ji2016wireless,golrezaei2014base}
to name a few. However, all these works make use of the simplistic
protocol model, where outage occurs if the intended receiver is at
a distance greater than a fixed distance from the transmitter or there
is another interfering transmitter present within the range of the
receiver. The other approach, which makes use of the physical model,
is where the outage occurs on the basis of the received signal-to-interference-and-noise
ratio (SINR).

Stochastic geometry has been widely applied to analyze the physical
layer metrics of the large scale wireless networks. For the case of
cache enabled cellular networks, the dynamics of content popularity,
propagation conditions and spatial locations are employed in \cite{PerabathiniBKDC15,sarzaidi2015,tamoor2015modeling,bastug2014cache,afshang2015fundamentals,chen2016cooperative,afshang2015modeling,AltieriPVG14}
to quantify the performance gains. The analysis of rate and energy
efficiency for single-tier cache enabled cellular networks is carried
out in \cite{bastug2014cache} and \cite{chen2016cooperative}. An
optimal content placement strategy is devised in \cite{chen2016cooperative},
which maximizes the rate coverage and the energy efficiency of the
single-tier cellular networks. The authors in \cite{afshang2015modeling}
and \cite{AltieriPVG14} consider clustered D2D networks, which operate
in isolation from the underlying cellular network. In \cite{afshang2015modeling},
the authors consider clustering to mimic spatial content locality
without explicitly considering content popularity and storage. It
is assumed that for a user in a given cluster, there will always be
a device in that cluster with the requested content i.e., a D2D wireless
link is always established. Whereas, the authors in \cite{AltieriPVG14}
consider D2D devices with caching and a Zipf type content popularity
distribution. Here, clustering is considered so that there are finite
transmissions within the cluster multiplexed in time as in TDMA and
one link is active at a given time. In \cite{afshang2015fundamentals},
the analysis is carried out with different D2D transmitter selection
schemes, but this work makes use of the same clustered users model
as in \cite{afshang2015modeling}. All these works do not take into
account the coexistence of D2D communication with cellular networks
and that if D2D communication is infeasible, the users can communicate
with the MBS.

For the case of multi-tier analysis with caching, the authors in \cite{sarzaidi2015}
and \cite{tamoor2015modeling} consider distributed caching, where
a user can access data from the caches of multiple small base stations
(SBSs) inside a cell. However, in case of the content availability
in any one of the SBSs within the cell, the user is always served
by its nearest SBS assuming that the content transfer takes place
among the SBSs. This is different from our case as we cannot expect
such level of cooperation between D2D helpers and need explicit characterization
of the distances of the individual D2D helpers from the arbitrary
user.

The selection of cellular and D2D modes is studied for the uplink
in \cite{lin2014spectrum} and \cite{elsawy2014analytical}. In \cite{lin2014spectrum},
the decision to transmit in D2D mode is based on the distance to the
receiver uniformly placed around the D2D transmitter, while in \cite{elsawy2014analytical},
it also depends on the distance from the BS. Both of these approaches
ignore the aspects of content availability, popularity and storage.

Various content replacement policies and storage techniques are studied
in \cite{niesen2014coded,dabirmoghaddam2014understanding,bacstuug2015transfer}.
\cite{dabirmoghaddam2014understanding} shows how updating a cache
by evicting the least recently used (LRU) content could provide performance
gains. It is shown that least frequently used (LFU) policy outperforms
LRU in \cite{fricker2012impact}. \cite{bacstuug2015transfer} explores
how caches could be updated by exploiting social ties between users
using transfer learning approach. \cite{niesen2014coded} proposes
coded caching for delay sensitive applications. In \cite{blaszczyszyn2015optimal}
and \cite{avrachenkov2016optimization}, the authors explore the effect
of geometric placement of caches to devise optimal content placement
strategies. A simpler, fixed-caching approach is adopted in \cite{bastug2014cache,LiuY15,AltieriPVG14}
and \cite{tamoor2015modeling}, where the cache is not updated and
the stored files are simply considered to follow the popularity distribution. 

The remainder of this paper is organized as follows: Section \ref{sec:System-Model}
describes the spatial setup, signal propagation, content popularity
and caching models, and the information-centric offloading paradigm
for both the NS and US schemes. Section \ref{sec:Distance-to-the}
provides the derivation of the distance between an arbitrary user
and its $i$th nearest D2D helper within the cell. The distribution
of this distance is then used to characterize the overall coverage
and the average rate for the NS and US schemes in Section \ref{sec:Link-Spectral-Efficiency}.
Section \ref{sec:Results-and-Performance} discusses the results and
validates our analysis with network simulations. Section \ref{sec:Conclusion}
concludes the paper.

\section{System Model\label{sec:System-Model}}

We consider a cellular downlink (DL) scenario of MBSs overlaid with
D2D helpers. The MBS schedules a requesting user with one of its neighboring
D2D helpers inside the cell if the helper has the requested file.
The network description and the key assumptions now follow.

\subsection{Spatial Model}

\begin{figure}
\begin{centering}
\includegraphics[scale=0.7]{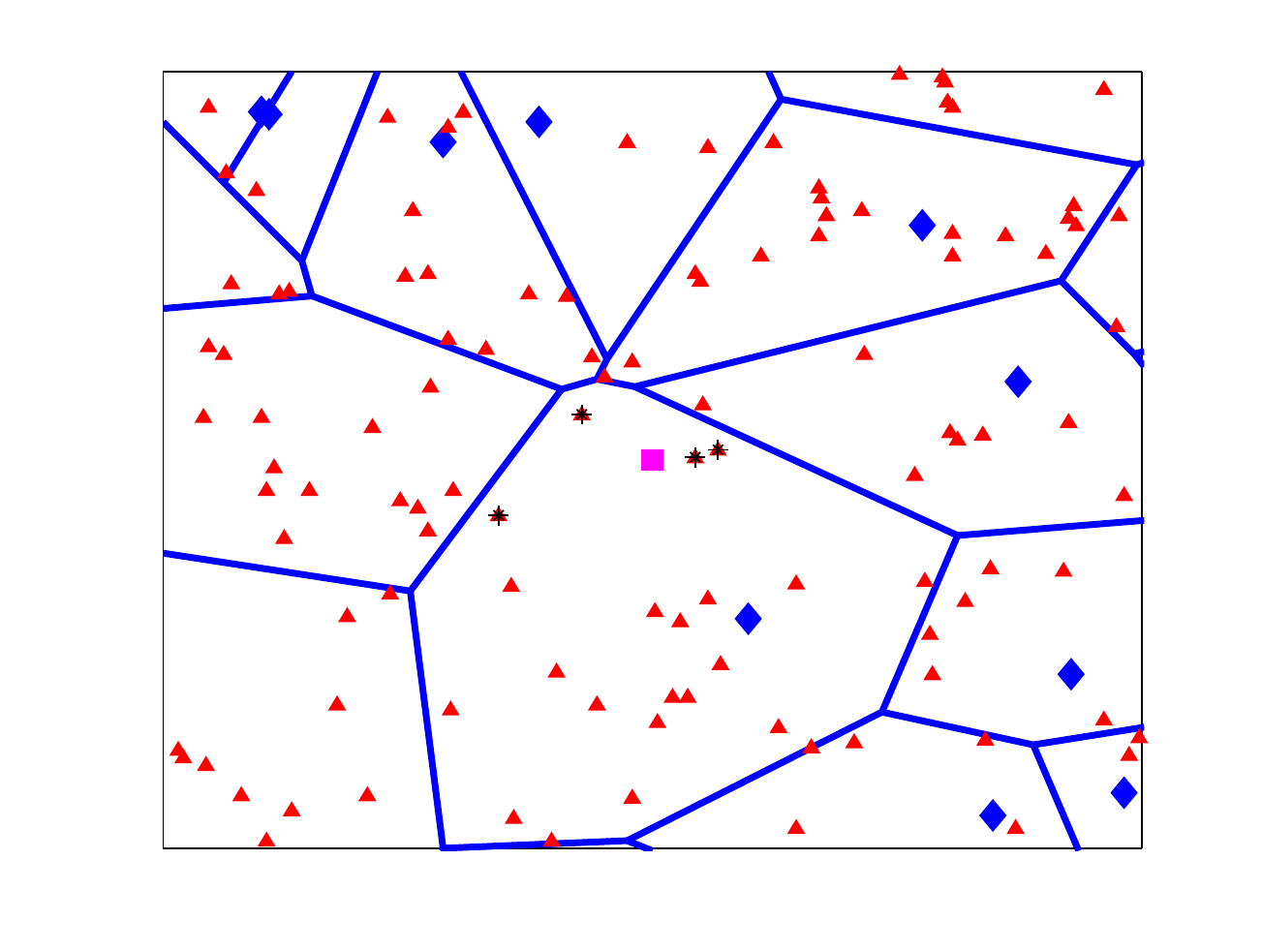}
\par\end{centering}
\centering{}\caption{Spatial model of the network. MBSs are depicted by blue, filled diamonds;
D2D helpers by red, filled triangles; and the requesting user by a
filled magenta square. The $k$ candidate D2D helpers for D2D communication
are marked with black asterisks (here, $k=4$) and $\lambda_{d}=10\lambda_{m}.$
\label{fig:SpatialModel}}
\end{figure}
According to the theory of HPPPs, the distribution of $\Phi\left(\mathcal{A}\right)$,
where $\mathcal{A}$ is a bounded borel set in $\mathbb{R}^{2}$,
is given as
\begin{equation}
\mathbb{P}[\Phi\left(\mathcal{A}\right)=n]=\frac{\left(\lambda\mu(\mathcal{A})\right)^{n}}{n!}\text{exp}\left(-\lambda\mu(\mathcal{A})\right),
\end{equation}
where $\lambda$ is the intensity of the HPPP and $\mu\left(\mathcal{A}\right)=\intop_{\mathcal{A}}dx$
is the Lebesgue measure on $\mathbb{R}^{2}$. For a disc of radius
$r$ in $\mathbb{R}^{2}$, $\mu\left(\mathcal{A}\right)=\pi r^{2}.$
We consider that the MBSs, D2D helpers and the requesting users are
distributed according to independent HPPPs $\Phi_{m}$, $\Phi_{d}$,
and $\Phi_{u}$ with intensities $\lambda_{m}$, $\lambda_{d}$ and
$\lambda_{u}$ respectively, where $\lambda_{u},\lambda_{d}\gg\lambda_{m}$.
The requesting users are associated to the nearest MBS and the user
association region is defined as
\begin{eqnarray}
\mathcal{S}_{i} & \overset{def}{=} & x\in\mathbb{R}^{2}\::\,\Vert y{}_{i}-x\Vert<\Vert y{}_{j}-x\Vert,\nonumber \\
 &  & \forall\,y{}_{i},y{}_{j}\in\Phi_{m},i\neq j,\label{eq:vcell}
\end{eqnarray}
where $\mathcal{S}_{i}$ represents a Voronoi cell of the MBS located
at $y_{i}\in\Phi_{m}$\footnote{We use the same notation to denote the the node itself and its distance
to the origin.}. Without any loss of generality, we measure performance at the requesting
user located at the origin. This follows from the palm distribution
of HPPPs and Slivnyak's theorem \cite{stoyan1987stochastic}. The
MBS selects one of the $k$ closest D2D helpers and establishes a
D2D link (the selection process is described in detail in the next
section). A realization of the spatial setup is shown in Fig. \ref{fig:SpatialModel}.

\subsection{Propagation Model and Spectrum Access}

We assume that both the cellular and D2D links experience channel
impairments including path loss and small-scale Rayleigh fading. The
power received at the origin from the MBS/ D2D helper located at $y\in\Phi_{n},n=\{m,d\}$
is given as $P_{n}h\,y^{-\alpha}\text{,}$ where%
\begin{comment}
 the power decay due to path-loss is modeled by $\Vert\textbf{y}-\textbf{x}\Vert^{-\alpha}$.
\end{comment}
{} $P_{m}$ and $P_{d}$ are the transmit powers of the MBS and D2D
helper respectively, $\alpha$ represents the path loss exponent ranging
between 2 and 5 and $h$ is the channel power. We assume that $h$
is a unit-mean exponential RV representing the squared-envelope of
Rayleigh fading. 

We consider an in-band overlay spectrum access strategy, where fixed
bandwidths $W_{m}$ and $W_{d}$ are allocated for cellular and D2D
communication respectively. We assume that there is universal frequency
reuse across the network, but the number of resource blocks is greater
than the number of users within the cell and hence, there is no intra-cell
interference. %
\begin{comment}

\subsection{Spectrum Access Model}

LTE standard defines DL bandwidth (BW) ranging from 1.25-20MHz. A
resource block (RB) is the smallest time-frequency slot that a MBS
can allocate to a user. It has the bandwidth of 180kHz and time duration
of 0.5 ms (half sub-frame) \cite{lte}. and instead of reserving a
particular BW or number of RBs for D2D communication, We propose to
reserve a fraction $\beta$ of the available DL resources for D2D
communication and the remaining $1-\beta$ for cellular communication.
\end{comment}

\subsection{Content Popularity and Caching Model \label{subsec:Content-Popularity-and}}

The performance of caching is crucially determined by the content
popularity distribution. It has been observed that the popularity
of data follows a Zipf popularity distribution where, the popularity
of the $c$th content is proportional to the inverse of $c^{\zeta}$
for some real, positive, skewness parameter $\zeta$. It is mathematically
represented as 
\begin{equation}
pop(c)=\rho c^{-\zeta}\;\;1\leq c\leq L,\label{eq:zipf}
\end{equation}
where $\rho=\left(\sum_{l=1}^{L}l^{-\zeta}\right)^{-1}$ is the distribution
normalizing factor and $L$ is the file library size. $\zeta=0$ corresponds
to uniform popularity while a higher value of $\zeta$ results in
a more skewed distribution. Empirical evidence shows that the value
for $\zeta$ exists between 0.6 to 0.8 for different content types
including web, file sharing, user generated content (UGC) and video
on demand (VoD) \cite{fricker2012impact}. We consider that the MBS
and the D2D helpers are equipped with caches of sizes $C_{m}$ and
$C_{d}$ respectively. All files are considered to have a unit size.
Our analysis can easily be extended for variable file sizes as each
memory slot will then contain a chunk of a file. We further assume
that user requests follow the independent reference model (IRM) as
introduced in \cite{fricker2012impact}. According to the IRM, the
user requests for a file in the library are independently generated
following the popularity distribution and there is no spatio-temporal
locality, i.e. identical contents have the same popularity in space
and time \cite{sarzaidi2015}. 

\subsubsection{Content Placement}

We consider that the MBS stores $C_{m}$ most popular files in its
cache. The MBS caches the most popular contents, which coincides with
the least frequently used (LFU) content placement strategy. Because
the content popularity does not rapidly change in time, LFU placement
is shown to be well-suited for the MBS. The cellular hit rate for
content $c$, which is the probability that the content $c$ is present
in the MBS's cache is given as
\begin{equation}
h_{m}(c)=\mathbb{I}_{c\leq C_{m}},
\end{equation}
where $\mathbb{I}_{c\leq C_{m}}$ is an indicator variable taking
the value 1 when $c=\{1,..,C_{m}\}$ and 0 otherwise.

When there is a set of candidate D2D helpers which can serve a single
user, the LFU placement for all D2D helpers is not optimal. Such a
scenario requires a collaborative content placement strategy which
takes into account the number and the locations of the D2D helpers
\cite{blaszczyszyn2015optimal,avrachenkov2016optimization}. Investigating
the optimal content placement strategy for D2D helpers for this network
setup is a research issue in itself and left for future work. We consider
a sub-optimal but tractable content placement strategy for the D2D
helpers to quantify the advantage of employing content-centric offloading
on D2D mode. We consider that each D2D helper stores the content $c$
in each memory slot independently according to the popularity distribution
$pop(c)$. The D2D hit rate for content $c$, is then given as 
\begin{eqnarray}
h_{d}(c) & = & 1-\mathbb{P}[\textnormal{Content }c\text{ not present in }C_{d}\text{ slots}]\nonumber \\
 & = & 1-[1-\rho c^{-\zeta}]^{C_{d}}.\label{eq:hitid}
\end{eqnarray}

\section{Information-Centric Offloading}

We assume that for a typical user requesting content $c$, the MBS
will examine the contents of up to $k$ neighboring D2D helpers within
the cell. The selection of the D2D helper depends on the helper selection
scheme and the popularity of the requested content itself. The following
proposition gives the probability for the selection of a particular
D2D helper. 
\begin{figure}
\begin{centering}
\subfloat[\label{fig:Mode-Selection-Probability}]{\begin{centering}
\includegraphics[scale=0.65]{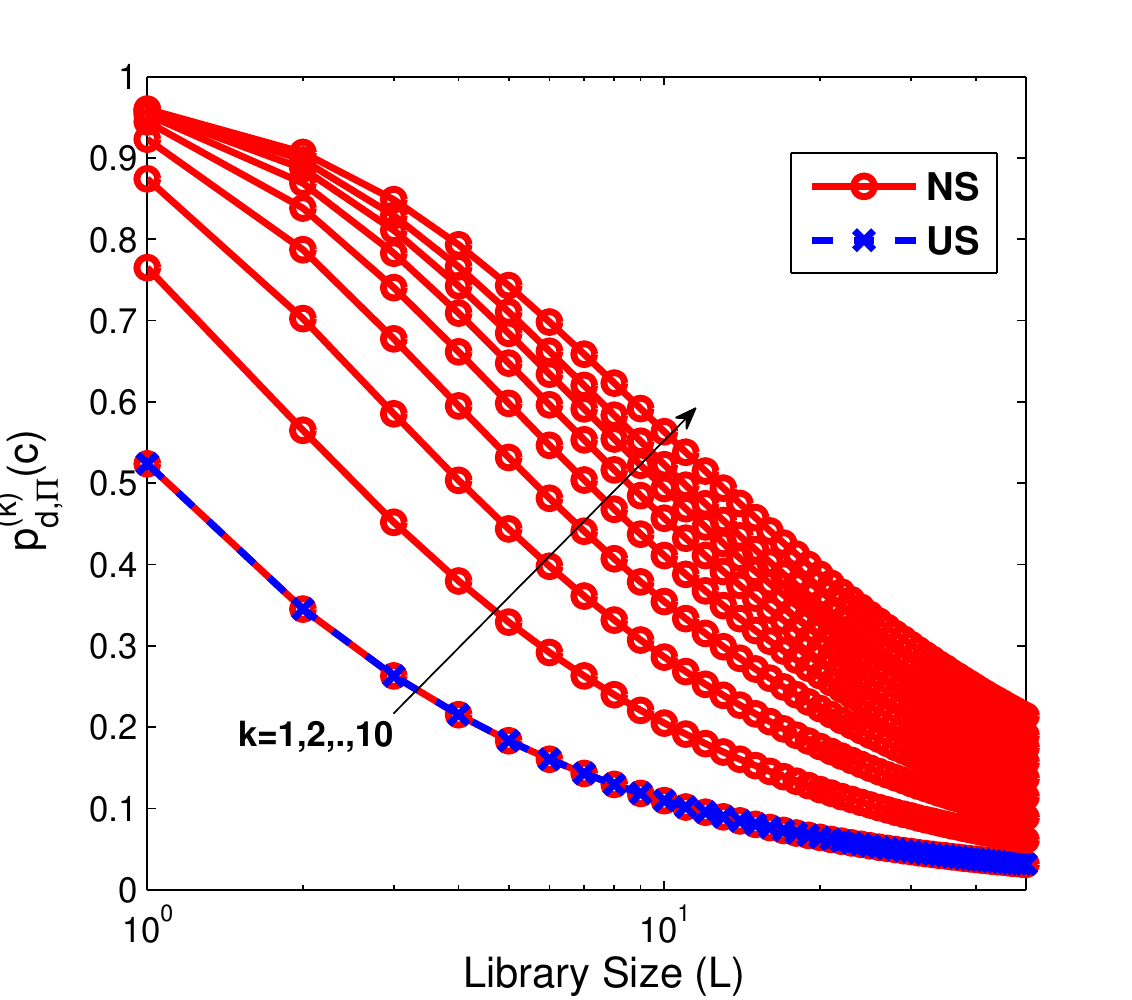}
\par\end{centering}
}\hfill{}\subfloat[\label{fig:Effect-of-}]{\centering{}\includegraphics[scale=0.65]{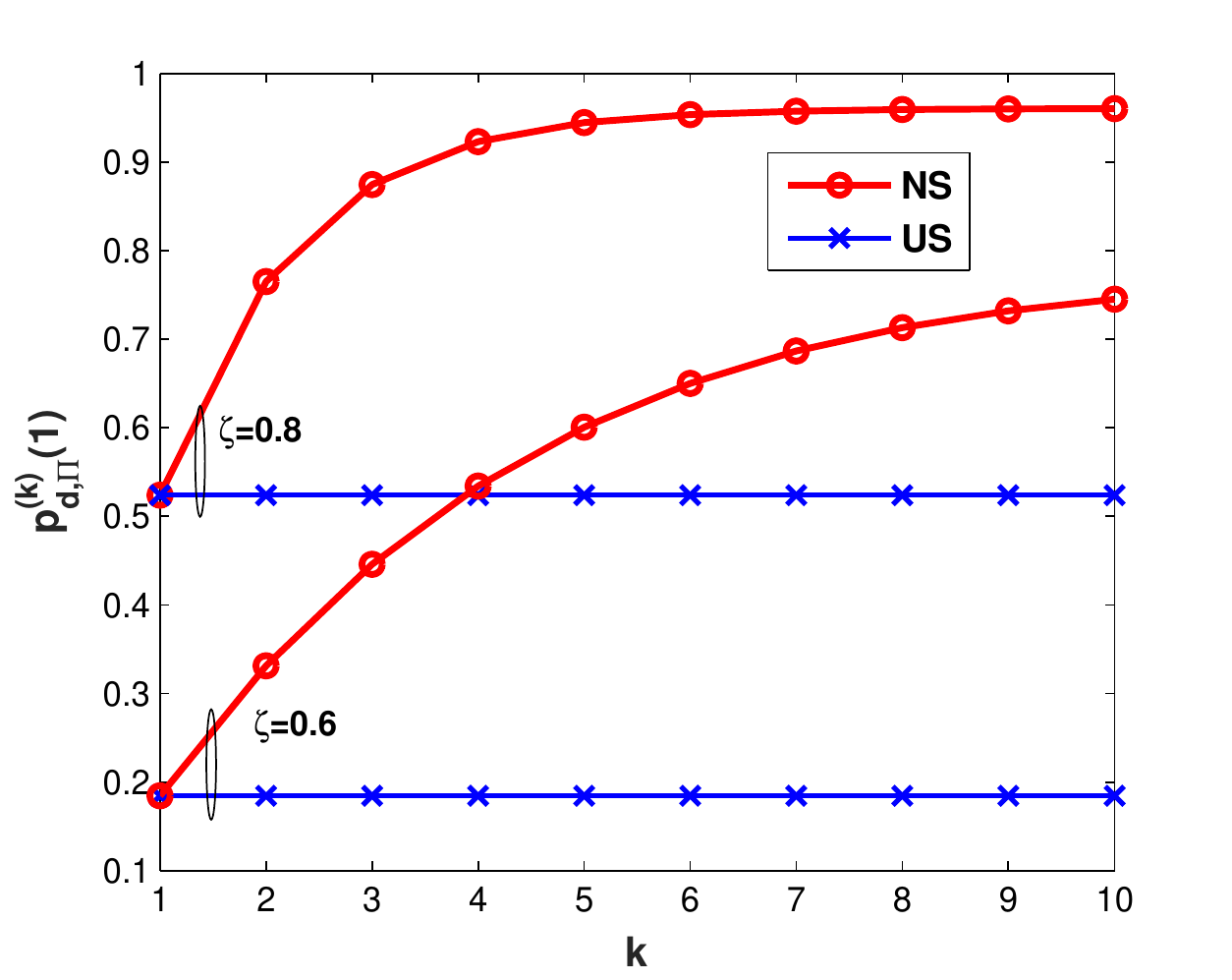}}
\par\end{centering}
\caption{Effect of helper selection schemes, requested contents, $k$ and $\zeta$
on the probability of D2D and cellular modes.}
\end{figure}

\begin{figure}
\centering{}\includegraphics[scale=0.65]{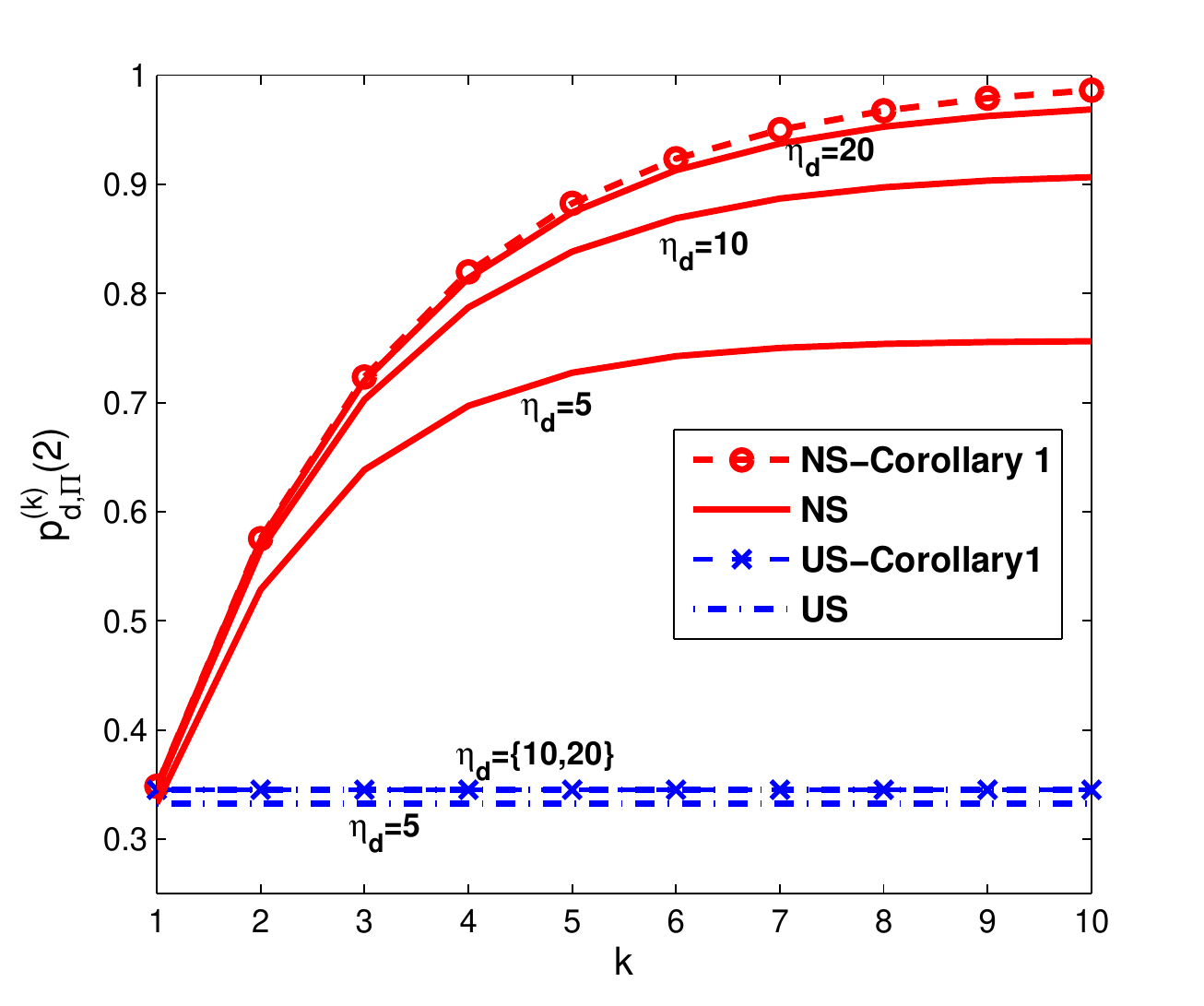}\caption{Effect of helper density on D2D mode probability.\label{fig:D2Dapprox}}
\end{figure}

\begin{prop}
The probability that an arbitrary user requesting content 'c' is served
by the ith nearest D2D helper within the cell under NS and US schemes
is given by 
\begin{equation}
p_{d,NS}^{(k)}(i,c)=\frac{343}{30}\sqrt{\frac{14}{\pi}}\frac{\textnormal{\ensuremath{\Gamma}}(i+3.5)\,\eta_{d}^{i}\,\tensor[_{2}]{F}{_{1}}\left(1,i+3.5;i+1;\frac{\eta_{d}}{(\eta_{d}+3.5)}\right)}{i!\,(\eta_{D}+3.5)^{i+3.5}}\left(1-h_{d}(c)\right)^{i-1}h_{d}(c),\label{eq:pd2diNS}
\end{equation}
and 
\begin{eqnarray}
p_{d,US}^{(k)}(i,c) & = & h_{d}(c)\biggl[\frac{343}{30}\sqrt{\frac{14}{\pi}}\frac{\textnormal{\ensuremath{\Gamma}}(k+4.5)\,\eta_{d}^{k+1}\,\tensor[_{2}]{F}{_{1}}\left(1,k+4.5;k+2;\frac{\eta_{d}}{(\eta_{d}+3.5)}\right)}{k(k+1)!\,(\eta_{d}+3.5)^{k+4.5}}+\label{eq:pd2diUS}\\
 &  & \sum_{j=0}^{k-i}\frac{3.5^{3.5}\textnormal{\ensuremath{\Gamma}}(i+j+3.5)\eta_{d}^{i+j}}{\textnormal{\ensuremath{\Gamma}}(3.5)\,(i+j)\,(i+j)!\,(\eta_{d}+3.5)^{i+j+3.5}}\biggr]\nonumber 
\end{eqnarray}
respectively, where $\eta_{d}=\lambda_{d}/\lambda_{m}$, $i=\{1,..k\},$
$\textnormal{\ensuremath{\Gamma}}(a)$ is the complete Gamma function
and $\tensor[_{2}]{F}{_{1}}\left(a,b;c;x\right)$ is the generalized
hypergeometric function.
\end{prop}
\begin{IEEEproof}
For the user to be served by the $i$th nearest D2D helper, there
must be at least $i$ D2D helpers inside the cell. In the NS scheme,
the user is served by the $i$th helper if no closer helper has the
requested content. This implies
\begin{eqnarray}
p_{d,NS}^{(k)}(i,c) & = & \mathbb{P}\left[N_{d}\geqslant i\right]\left(1-h_{d}(c)\right)^{i-1}h_{d}(c),\label{eq:pd2di}
\end{eqnarray}
where $N_{d}$ is the number of D2D helpers in a Voronoi cell whose
probability mass function is given as \cite{yu2013downlink}
\begin{equation}
\mathbb{P}\left[N_{d}=j\right]=\frac{3.5^{3.5}\textnormal{\ensuremath{\Gamma}}(j+3.5)(\lambda_{d}/\lambda_{m})^{j}}{\textnormal{\ensuremath{\Gamma}}(3.5)\,i!\,(\lambda_{d}/\lambda_{m}+3.5)^{i+3.5}}.
\end{equation}
This implies $\mathbb{P}\left[N_{d}\geqslant i\right]=1-\sum_{j=0}^{i-1}\mathbb{P}\left[N_{d}=j\right]$.
Substituting this expression in (\ref{eq:pd2di}) gives (\ref{eq:pd2diNS}). 

For the US scheme, the user is served by the $i$th helper if it is
uniformly selected and has the requested content. This implies
\begin{equation}
p_{d,US}^{(k)}(i,c)=h_{d}(c)\left[\frac{1}{k}\mathbb{P}\left[N_{d}>k\right]+\sum_{j=0}^{k-i}\frac{1}{i+j}\mathbb{P}\left[N_{d}=i+j\right]\right],\label{eq:pdUSi}
\end{equation}
Substituting the expressions for $\mathbb{P}\left[N_{d}>k\right]$
and $\mathbb{P}\left[N_{d}=i+j\right]$ completes the proof.
\end{IEEEproof}
The probability that an arbitrary user is served in D2D mode under
NS and US schemes is a straightforward summation of $p_{d,NS}^{(k)}(i,c)$
and $p_{d,US}^{(k)}(i,c)$ over $i=\{1,2..,k\}.$ This gives 
\begin{equation}
p_{d,NS}^{(k)}(c)=\sum_{i=1}^{k}p_{d,NS}^{(k)}(i,c),\label{eq:pd2dNSc}
\end{equation}
and
\begin{equation}
p_{d,US}^{(k)}(c)=p_{d,US}^{(1)}(c)=h_{d}(c)\left[1-(1+3.5^{-1}\eta_{d})^{-3.5}\right].\label{eq:pd2dUSc}
\end{equation}
It is interesting to note that in case of the US scheme, $p_{d,US}^{(k)}(c)$
is independent of $k$. This is because the probability of the D2D
mode depends on the contents of only one helper selected at random. 
\begin{cor}
\label{cor:hitSimple} The bounds on $p_{d,\varPi}^{(k)}(i,c),\varPi=\{NS,US\}$
with respect to $\eta_{d}$ are given as
\begin{equation}
p_{d,NS}^{(k)}(i,c)\leq\left(1-h_{d}(c)\right)^{i-1}h_{d}(c)\label{eq:pdiNSsimple}
\end{equation}
\begin{comment}
\begin{equation}
p_{d,NS}^{(k)}(c)\leq1-[1-\rho c^{-\zeta}]^{k\,C_{d}}\label{eq:pdNSsimple}
\end{equation}
\end{comment}
and
\begin{equation}
p_{d,US}^{(k)}(i,c)\leq\frac{1}{k}h_{d}(c)\label{eq:pdiUSsimple}
\end{equation}
\begin{comment}
\begin{equation}
p_{d,US}^{(k)}(c)\leq1-[1-\rho c^{-\zeta}]^{C_{d}},\label{eq:pdUSsimple}
\end{equation}
\end{comment}
{} where the equalities hold when $\lambda_{d}\gg\lambda_{m}$ and $\eta_{d}\rightarrow\infty$.
\end{cor}
\begin{IEEEproof}
As $\eta_{d}\rightarrow\infty$, $\mathbb{P}\left[N_{d}\geqslant k\right]\rightarrow1$,
i.e. there are definitely at least $k$ D2D helpers within the cell
and (\ref{eq:pd2di}) reduces to $\left(1-h_{d}(c)\right)^{i-1}h_{d}(c)$.
It can be easily seen from (\ref{eq:pdUSi}) that as $\mathbb{P}\left[N_{d}\geqslant k\right]\rightarrow1$,
$\mathbb{P}\left[N_{d}=i+j\right]\rightarrow0,$ where $i+j<k$. Hence,
$p_{d,US}^{(k)}(c)$ reduces to $h_{d}(c)$.
\end{IEEEproof}
Before moving on to further analysis, we explore the behavior of the
D2D mode probability in Figs. \ref{fig:Mode-Selection-Probability}
and \ref{fig:Effect-of-}. The values of the simulation parameters
used in plotting the results are listed further on in Table \ref{tab:List}
unless stated otherwise. We can see that there is a rapid increase
in $p_{d,NS}^{(k)}(c)$ initially with the increase in $k$, but diminishing
gains are observed when $k$ is further increased. A sharp decrease
in $p_{d,NS}^{(k)}(c)$ is observed as the requested content becomes
less popular or the skewness parameter $\zeta$ decreases. As established
in (\ref{eq:pd2dUSc}), we see that there is no effect of increasing
$k$ on $p_{d,US}^{(k)}(c).$ In Fig. \ref{fig:D2Dapprox}, we plot
the D2D mode probabilities $p_{d,\varPi}^{(k)}(c)=\sum_{i=1}^{k}p_{d,\varPi}^{(k)}(i,c),\varPi=\{NS,US\}$
using the upper bounds from Corollary \ref{cor:hitSimple} and compare
them for the actual values of $p_{d,\varPi}^{(k)}(c)$ in (\ref{eq:pd2dNSc})
and (\ref{eq:pd2dUSc}). We see that the deviation for the NS scheme
becomes large as the value of $k$ increases, but convergence is fast
and the bounds are fairly tight for $\eta_{d}\geq10$. 

\section{Distance to the $i$th nearest D2D helper within a Macrocell\label{sec:Distance-to-the}}

One of the main contributions of this paper is to characterize the
distribution of the distance between the typical user and the $i$th
nearest D2D helper within the macro cell. It is a well-known fact
that the distance between the nearest neighbors for a 2-D Poisson
process is Rayleigh distributed and this has been widely adopted for
the stochastic geometry analysis of cellular networks \cite{lin2014spectrum,elsawy2014analytical},
\cite{andrews2011tractable,bastug2014cache}. In our case, however,
the MBS only keeps a record of the files stored in the memory of D2D
helpers within its coverage region. Therefore, it can only connect
the requesting user with the helpers within its cell. Fig. \ref{fig:SpatialModel}
illustrates that in our spatial setup, the $i$th nearest D2D helper
is not always within the macrocell. Hence, this adds a layer of complexity
to our model as the distance is no longer independent of the geometrical
attributes of the cell, including its shape and size. 

\begin{figure}
\begin{centering}
\includegraphics[scale=0.65]{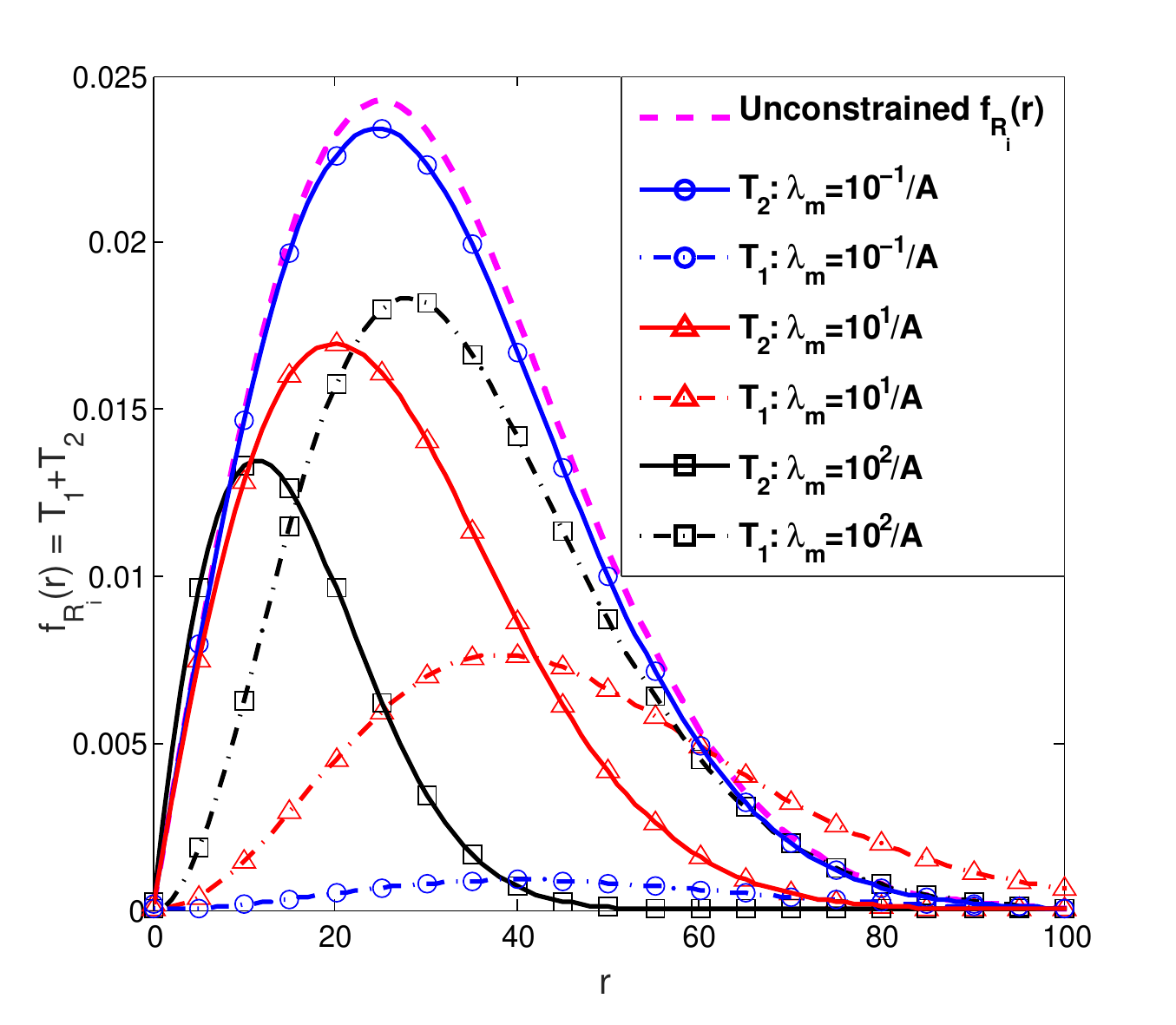}
\par\end{centering}
\caption{Effect of varying $\lambda_{m}$ on $T_{1}$ and $T_{2}$: $i=1,\lambda_{d}=200/A,A=\pi500^{2}.$
\label{fig:T1T2}}
\end{figure}

The distribution of the exact shape and size of a typical Voronoi
cell in a 2-D Poisson Voronoi tessellation is still unknown. In their
analysis of bivariate Poisson processes in \cite{foss1996voronoi},
Foss and Zuyev make use of the maximal disk approximation for the
Voronoi cell. The maximal disk, $B_{max}$, is the largest disk centered
at the MBS inscribing the Voronoi cell. The exact characterization
of the distribution of the radius $X$ of $B_{max}$ is straight forward
as it is the probability that there is no other BS at a distance $2x$
from the tagged BS and is expressed as $\mathbb{P}[X\geq2x]=\text{exp}\left(-4\lambda_{m}\pi x^{2}\right).$
This implies

\begin{equation}
f_{X}(x)=8\lambda_{m}\pi x\text{ exp}(-4\lambda_{m}\pi x^{2}).\label{eq:rhomax}
\end{equation}
In this work, we utilize the maximal ball approximation for the Voronoi
cell to derive the distribution of the distance between the typical
user and its $i$th nearest D2D helper\footnote{In our previous work in \cite{aafzal2016}, we show that the maximal
disk approximation is not accurate when the user is placed at a fixed
distance from the MBS. In fact, a disk with the same area as the Voronoi
cell better approximates the distance. However, in this case, we will
show that when the distance $Y$ between the requesting user and the
MBS is random, the inscribed disk approximation accurately approximates
the distance.}. The following Lemmas provide some useful preliminary results which
are necessary conditions for the characterization of the distance
distribution. 
\begin{lem}
The probability that the typical user lies inside $B_{max}$ is the
probability that the radius of $B_{max}$ is greater than the distance
between the MBS and typical user. It is a constant value and is equal
to 
\begin{equation}
p_{in}=\mathbb{P}\left[X\geq Y\right]=1/5,\label{eq:pin}
\end{equation}
where $f_{Y}(y)=2\lambda_{m}\pi y\,\textnormal{exp}(-\lambda_{m}\pi y^{2}).$
\end{lem}
\begin{IEEEproof}
The distance between the typical user and its tagged MBS is Rayleigh
distributed \cite{andrews2011tractable}. This implies $p_{in}=\int_{0}^{\infty}\left[1-F_{X}(y)\right]f_{Y}(y)\,dy$,
where $F_{X}(y)=\int_{0}^{y}f_{X}(x)\,dx$. Solving the integrals
we get (\ref{eq:pin}). 
\end{IEEEproof}
\begin{lem}
The probability that there are at least i D2D helpers inside $B_{max}$
is given as 
\begin{equation}
p_{N_{d}}^{(i)}=5(1+4\eta_{d}^{-1})^{-i}+\frac{10}{3}(1+6\eta_{d}^{-1})^{-i}-8(1+5\eta_{d}^{-1})^{-i}.\label{eq:pNdi}
\end{equation}
\end{lem}
\begin{IEEEproof}
Given a disk $B_{max}$ with radius $X=x$, the probability that there
are at least $i$ D2D helpers inside the disk is given as
\begin{eqnarray*}
\mathbb{P}\left[N_{d}\geq i|X=x\right] & = & 1-\sum_{j=0}^{i-1}\frac{\left(\lambda_{d}\pi x^{2}\right)^{j}}{j!}\textnormal{exp}\left(-\lambda_{d}\pi x^{2}\right)=1-\frac{\textnormal{\ensuremath{\Gamma}}(i,\lambda_{d}\pi x^{2})}{\textnormal{\ensuremath{\Gamma}}(i)}.
\end{eqnarray*}
Integrating over $X=x$ while ensuring $X>D$, we obtain 
\begin{equation}
p_{N_{d}}^{(i)}=\intop_{0}^{\infty}\biggl[\intop_{y}^{\infty}\left[1-\frac{\textnormal{\ensuremath{\Gamma}}(i,\lambda_{d}\pi x^{2})}{\textnormal{\ensuremath{\Gamma}}(i)}\right]\,f_{X}(x)\,F_{Y}(x)\,dx\biggr]\,f_{Y}(y)\,dy,
\end{equation}
where $F_{Y}(x)=\int_{0}^{x}f_{Y}(y)\,dy$. Simplification after evaluating
the integrals and normalizing with $p_{in}$ yields (\ref{eq:pNdi}).
\end{IEEEproof}
We now give the distribution of distance in the following Theorem. 
\begin{thm}
\label{prop:dist}The distribution of the distance between the typical
user and its ith nearest D2D helper within the cell can be well approximated
using the inscribed disk approximation for a Voronoi cell and is given
as
\begin{multline}
f_{R_{i}}(r)=\biggl[\intop_{0}^{\infty}f_{Y}(y)\intop_{a_{1}}^{a_{2}}f_{i,1}(r,y,x)\,f_{X}(x)\,F_{Y}(x)dx\,dy+f_{i,2}(r)\kappa(r)\biggr]\frac{1}{p_{in}\,p_{N_{d}}^{(i)}},\label{eq:fr}
\end{multline}
where, 
\begin{equation}
f_{i,1}(r)=\frac{\lambda_{d}^{i}}{\textnormal{\ensuremath{\Gamma}}(i)}\nabla(r,y,x)^{i-1}\nabla^{'}(r,y,x)\,\textnormal{exp}(-\lambda_{d}\nabla(r,y,x)),\label{eq:f1r}
\end{equation}
\begin{equation}
f_{i,2}(r)=2\frac{\left(\lambda_{d}\pi\right)^{i}}{\textnormal{\ensuremath{\Gamma}}(i)}r^{2i-1}\textnormal{exp}(-\lambda_{d}\pi r^{2}),\label{eq:f2r}
\end{equation}
\begin{eqnarray}
\kappa(r) & = & \frac{\textnormal{exp}\left(-4b\right)}{15}+b\sqrt{\pi}\biggl[\frac{\sqrt{6}}{9}\textnormal{exp}\left(\frac{b}{6}\right)\textnormal{erfc}\left(\frac{5\sqrt{6b}}{6}\right)-\frac{4\sqrt{5}}{25}\textnormal{exp}\left(-\frac{4b}{5}\right)\textnormal{erfc}\left(\frac{4\sqrt{5b}}{5}\right)\biggr],\nonumber \\
\label{eq:kappaR}
\end{eqnarray}
where $b=\lambda_{m}\pi r^{2}$, $\nabla(r,y,x)=r^{2}\arccos\left(\frac{\omega_{1}}{2y\,r}\right)+x^{2}\arccos\left(\frac{\omega_{2}}{2y\,x}\right)-\frac{1}{2}\sqrt{4y^{2}x^{2}-\omega_{2}^{2}}$,
$\omega_{1}=r^{2}+y^{2}-x^{2}$, $\omega_{2}=x^{2}+y^{2}-r^{2}$,
$\nabla^{'}(r,y,x)$ is the derivative of $\nabla(r,y,x)$ with respect
to $r$, $a_{1}=\max(y,r-y)$ and $a_{2}=r+y$. 
\end{thm}
\begin{IEEEproof}
Please refer to Appendix \ref{sec:Proof1}.
\end{IEEEproof}
The expression in (\ref{eq:fr}) is validated with network simulations
in Section \ref{sec:Results-and-Performance}. Before further analysis,
we develop some insights on the derived distance distribution in (\ref{eq:fr}).
We can write $f_{R_{i}}=T_{1}+T_{2}$, where $T_{1}=(p_{in}\,p_{N_{d}}^{(i)})^{-1}\intop_{0}^{\infty}\intop_{a_{1}}^{a_{2}}f_{i,1}(r,y,x)\,f_{X}(x)\,F_{Y}(x)dx\,f_{Y}(y)dy$
and $T_{2}=(p_{in}\,p_{N_{d}}^{(i)})^{-1}f_{i,2}(r)\kappa(r)$. We
wish to see how the density of MBSs impacts $T_{1}$ and $T_{2}$
and in turn $f_{R_{i}}$.
\begin{cor}
\label{cor:sparseDist}For sparse networks, i.e. when $\lambda_{m}\rightarrow0(\eta_{d}\rightarrow\infty),$
$f_{R_{i}}(r)$ reduces to the distribution of distance to the unconstrained
nearest D2D helper and is given by (\ref{eq:f2r}).
\end{cor}
\begin{IEEEproof}
Referring to Appendix \ref{sec:Proof1}, we see that when $\lambda_{m}\rightarrow0$,
$x\gg r$ and $b(o,r)$ almost surely lies inside $B_{max}$. This
in turn means $T_{1}\rightarrow0$. However, as $\lambda_{m}\rightarrow0$,
we see from (\ref{eq:kappaR}) and (\ref{eq:pNdi}) that $\kappa(r)=1/15$
and $p_{N_{d}}^{(i)}=1/3$. As $p_{in}=1/5$ is fixed, $T_{2}$ reduces
to $f_{i,2}(r)$.
\end{IEEEproof}
Fig. \ref{fig:T1T2} reinforces the result in Cor. \ref{cor:sparseDist}.
We compare (\ref{eq:fr}) with the unconstrained $i$th nearest neighbor
distribution \cite{moltchanov2012survey}. We see that when the network
is sparse, the term $T_{2}$ dominates $f_{R_{i}}(r)$ and the distribution
of the distance to the $i$th nearest neighbor essentially approaches
that of unconstrained case. This conclusion can be intuitively explained
as we would expect that for very large cell sizes, the $i$th nearest
D2D helper will reside in the same macrocell. However, as the MBS
density increases, $T_{1}$ begins to increase and cannot be ignored.
That is when (\ref{eq:fr}) begins to significantly deviate from (\ref{eq:f2r}).

\section{Performance Analysis under NS and US schemes\label{sec:Link-Spectral-Efficiency}}

To assess the performance of cellular networks enhanced with coordinated
D2D communication, we define the following quality-of-service (QoS)
parameters. 

\subsection{Overall coverage probability }

The typical user is in coverage in cellular mode when the received
signal to interference and noise ratio (SINR) is greater than a certain
modulation dependent decoding threshold $\tau_{m}$. This is mathematically
characterized as 
\[
\varGamma_{m}=\mathbb{P}\left[SINR_{m}\geq\tau_{m}\right].
\]
Similarly, in D2D mode, when the user is served by its $i$th nearest
D2D helper, the coverage probability is written as 
\[
\Gamma_{d,i}=\mathbb{P}\left[SINR_{d,i}\geq\tau_{d}\right],
\]
where $\tau_{d}$ is the SINR threshold in D2D mode. We define the
overall coverage probability $\Gamma_{\varPi}^{(k)}(c)$ of an arbitrary
user requesting content $c$ by the following expression
\begin{equation}
\Gamma_{\varPi}^{(k)}(c)=(1-p_{d,\varPi}^{(k)}(c))\Gamma_{m}+p_{d,\varPi}^{(k)}(c)\Gamma_{d,\varPi}^{(k)}(c),\label{eq:OverallCov}
\end{equation}
where $\Gamma_{d,\varPi}^{(k)}(c)$ is the probability of coverage
in D2D mode for a given $k$, content request $c$ and D2D helper
selection scheme $\varPi=\{NS,US\}$. Here, $p_{d,\varPi}^{(k)}(c)$
and $\left(1-p_{d,\varPi}^{(k)}(c)\right)$ are the probabilities
for D2D and cellular modes respectively. 

Our first step is to obtain the D2D and cellular coverage probabilities
of a typical link. When the typical user is operating in cellular
mode, we have from \cite{andrews2011tractable}
\begin{equation}
\Gamma_{m}=\pi\lambda_{m}\int_{0}^{\infty}\textnormal{exp}\left(-\pi\lambda_{m}\nu\left(1+\delta_{m}(s_{m},\alpha)\right)-\frac{\tau_{m}\sigma^{2}}{P_{m}}\nu^{\alpha/2}\right),\label{eq:covC}
\end{equation}
where $\Gamma_{m}$ is the cellular coverage probability, $s_{m}=\tau_{m}\nu^{\alpha}$
and $\delta_{m}(s_{m},\alpha)=\tau_{m}^{2/\alpha}\intop_{\tau_{m}^{-2/\alpha}}(1+u^{\alpha/2})du$.

\begin{prop}
\label{prop:covD}Given that the typical user is served in D2D mode,
the probability of coverage under the NS and US schemes can be expressed
as
\begin{equation}
\Gamma_{d,\varPi}^{(k)}(c)=\frac{\sum_{i=1}^{k}p_{d,\varPi}^{(k)}(i,c)\Gamma_{d,i}}{\sum_{i=1}^{k}p_{d,\varPi}^{(k)}(i,c)},\;\varPi=\{NS,US\},\label{eq:covDk}
\end{equation}
where $\Gamma_{d,i}$ is the coverage probability when the typical
user is served by the ith nearest D2D helper and is given as 

\begin{eqnarray}
\varGamma_{d,i} & \approx & \intop_{r=0}^{\infty}\textnormal{exp}\left(-2\pi\frac{s_{d}\tilde{\lambda}_{m}\delta_{d}(s_{d},\alpha)}{(\alpha-2)}\right)\textnormal{exp}\left(-s_{d}\frac{\sigma^{2}}{P_{d}}\right)f_{R_{i}}(r)\,dr\label{eq:covD2D}
\end{eqnarray}
where $s_{d}=\tau_{d}r^{\alpha}$, $\delta_{D}(s,\alpha)=\mathbb{E}_{Q}\left[q^{-(\alpha-2)}\tensor[_{2}]{F}{_{1}}\left(1,\bar{\alpha};1+\bar{\alpha};-sq^{-\alpha}\right)\right]$,
$\bar{\alpha}=1-2/\alpha$ and $f_{Q}(q)=2\pi\tilde{\lambda}_{M}q\,\textnormal{\textnormal{\textnormal{exp}}}(-\tilde{\lambda}_{M}\pi q^{2}).$
\end{prop}
\begin{comment}
\begin{prop}
\begin{equation}
\mathbb{E}_{q}\left[\text{exp}\left(-\pi\lambda_{m}s^{2/\alpha_{d}}\intop_{u_{min}}^{\infty}\frac{1}{1+u^{\alpha_{d/2}}}du\right)\right]\label{eq:}
\end{equation}

with $u_{min}=\left(\frac{q}{r\tau^{1/\alpha_{d}}}\right)^{2}$.
\end{prop}
\begin{IEEEproof}
After applying a change of variables to (\ref{eq:laplaceD2D})-$(c)$,
we obtain (\ref{eq:}).
\end{IEEEproof}
\end{comment}

\begin{IEEEproof}
Please refer to Appendix \ref{sec:Proof2}
\end{IEEEproof}

\subsection{Average Rate}

Using a similar exposition as in the previous subsection, we express
the average rate $T_{\varPi}^{(k)}(c)$ experienced by an arbitrary
user requesting content $c$ under the NS and US schemes as 
\begin{equation}
T_{\varPi}^{(k)}(c)=\left(1-p_{d,\varPi}^{(k)}(c)\right)\overline{R_{m,\varPi}}(c)+p_{d,\varPi}^{(k)}(c)\overline{R_{d,\varPi}}(c)\text{\;bps},\label{eq:tp}
\end{equation}
where $p_{d,\varPi}^{(k)}(c)$ and $\left(1-p_{d,\varPi}^{(k)}(c)\right)$
are the probabilities for D2D and cellular communication and $\overline{R_{m,\varPi}}(c)$
and $\overline{R_{d,\varPi}}(c)$ are respectively the average cellular
and D2D rates. 
\begin{prop}
\label{prop:Rd}The average rate experienced by an arbitrary user
requesting content 'c' in D2D mode under NS and US schemes is expressed
as 
\begin{equation}
\overline{R_{d,\varPi}^{(k)}}(c)=W_{d}\gamma\left(\eta_{u_{d,\varPi}}\right)\frac{\sum_{i=1}^{k}p_{d,\varPi}^{(k)}(i,c)R_{d,i}}{p_{d,\varPi}^{(k)}(c)},
\end{equation}
\begin{equation}
R_{d,i}=\mathbb{E}\left[\textnormal{log}_{2}(1+SINR_{d,i})\right],\label{eq:Rdi}
\end{equation}
where $\varPi=\{NS,US\}$, $\gamma(a)=\left(1-\textnormal{exp}\left(-a\right)\right)/a$,
$W_{d}$ is the bandwidth reserved for D2D communication and $\eta_{u_{d,\varPi}}=\lambda_{u_{d,\varPi}}/\lambda_{m}$.
Here, $\lambda_{u_{d,\varPi}}=\lambda_{u}\rho\sum_{c=1}^{L}c^{-\zeta}p_{d,\varPi}^{(k)}(c)$
is the average density of users operating in D2D mode. 
\end{prop}
\begin{IEEEproof}
The average D2D rate can be written as
\begin{equation}
\overline{R_{d,\varPi}^{(k)}}(c)=\mathbb{E}\left[\frac{W_{d}}{N_{u_{d,\varPi}}}\right]R_{d,\varPi}^{(k)}(c),
\end{equation}
where $R_{d,\varPi}^{(k)}(c)$ is the\textcolor{black}{{} average capacity
of the link when content $c$ is requested. It is obtained by taking
the expectation of (\ref{eq:Rdi}), which is the average ergodic capacity
of a D2D link between an arbitrary user and its $i$th nearest D2D
helper.} $\mathbb{E}\left[W_{d}/N_{u_{d,\varPi}}\right]$ is the average
bandwidth available for the communication on a D2D link where $N_{u_{d,\varPi}}\sim\textnormal{Poisson }(\eta_{u_{d,\varPi}})$
is the number of simultaneously active users in D2D mode inside the
cell\footnote{To simplify the analysis, we assume that the number of users in D2D
mode is the same as the number of active D2D helpers. This might not
be the case in reality as one helper can serve multiple users in its
vicinity if they request the same file. With our analysis, such a
situation will translate into the transmission of the same file by
the same D2D helper but on a separate portion of the spectrum. }. Hence,

\begin{eqnarray}
\mathbb{E}\left[1/N_{u_{d,\varPi}}\right] & = & \sum_{n=1}^{\infty}\frac{1}{n}\frac{(\eta_{u_{d,\varPi}})^{n-1}}{(n-1)!}\textnormal{exp}\left(-\eta_{u_{d,\varPi}}\right),\nonumber \\
 & = & \eta_{u_{d,\varPi}}^{-1}\left(1-\textnormal{exp}\left(-\eta_{u_{d,\varPi}}\right)\right).
\end{eqnarray}
where the summation starts from $n=1$ to account for the presence
of the user under consideration. 
\end{IEEEproof}
\begin{prop}
\label{prop:Rc}The average rate experienced by an arbitrary user
requesting content 'c' in cellular mode under NS and US schemes is
expressed as
\begin{equation}
\overline{R_{m,\varPi}^{(k)}}(c)=W_{m}\hat{R}_{m}\gamma\left(\eta_{u_{m,\varPi}}\right)
\end{equation}
where $\hat{R}_{m}(c)=R_{m}\left(\mathbb{I}_{c\leq C_{m}}+\beta\,\mathbb{I}_{c>C_{m}}\right)$
is the average capacity of cellular links, $\mathbb{I}_{c\leq C_{m}}$
and $\mathbb{I}_{c>C_{m}}$ are respectively the cellular hit and
miss rates, $R_{m}=\mathbb{E}\left[\textnormal{log}_{2}(1+SINR_{m})\right]$
, $\beta$ is the backhaul delay coefficienct, $W_{m}$ is the cellular
bandwidth, and $\eta_{u_{m,\varPi}}=\lambda_{u_{m,\varPi}}/\lambda_{m}$
with $\lambda_{u_{m,\varPi}}=\lambda_{u}(1-\rho\sum_{c=1}^{L}c^{-\zeta}p_{d,\varPi}^{(k)}(c))$.
\end{prop}
\begin{IEEEproof}
The proof is on the similar lines as Proposition \ref{prop:Rd} with
the exception that the average density of cellular users is $\lambda_{u}-\lambda_{u_{d,\varPi}}$,
i.e. all the users which are not operating in D2D mode are shifted
to cellular mode. Furthermore, when the requested content is not present
in the MBS cache, the average rate of the cellular link is reduced
by a factor $\beta$, which accounts for the delay introduced by fetching
the content from backhaul.
\end{IEEEproof}

\section{Results and Discussion\label{sec:Results-and-Performance}}

\begin{table}[t]
\begin{centering}
\begin{tabular}{|l|l|l|}
\hline 
Parameter & Description & Value\tabularnewline
\hline 
\hline 
$\alpha$ & Path loss exponent & 4\tabularnewline
\hline 
$\lambda_{m},\lambda_{d},\lambda_{u}$ & MBS, D2D helper and user density & $[10,100,200]/\pi500^{2}$\tabularnewline
\hline 
$\zeta$ & Popularity skewness parameter & 0.8\tabularnewline
\hline 
$c,L$ & Requested content, Library size & 1, $10^{4}$\tabularnewline
\hline 
$\beta$ & Backhaul delay coefficient & 0.8\tabularnewline
\hline 
$C_{d},C_{m}$ & D2D and MBS cache sizes & 20, 500\tabularnewline
\hline 
$W_{m},W_{d}$ & Cellular and D2D bandwidth & $[7,3]$ MHz \tabularnewline
\hline 
$P_{m},P_{d}$ & Cellular and D2D transmit power & $[30,23]$ dBm\tabularnewline
\hline 
$\tau_{m},\tau_{d}$ & Cellular and D2D SINR threshold & $[30,30]$ dBm\tabularnewline
\hline 
$\sigma^{2}$ & Noise power & -110 dBm\tabularnewline
\hline 
\end{tabular}
\par\end{centering}
\caption{List of simulation parameters\label{tab:List}}
\end{table}

In this section, we will give some key results and verify our analysis
with Monte Carlo simulations. For our simulation setup, the MBSs and
D2D helpers are distributed according to HPPPs with intensities $\lambda_{m}$
and $\lambda_{d}$ respectively and the performance is measured at
the origin. 

\begin{figure}
\centering{}\includegraphics[scale=0.6]{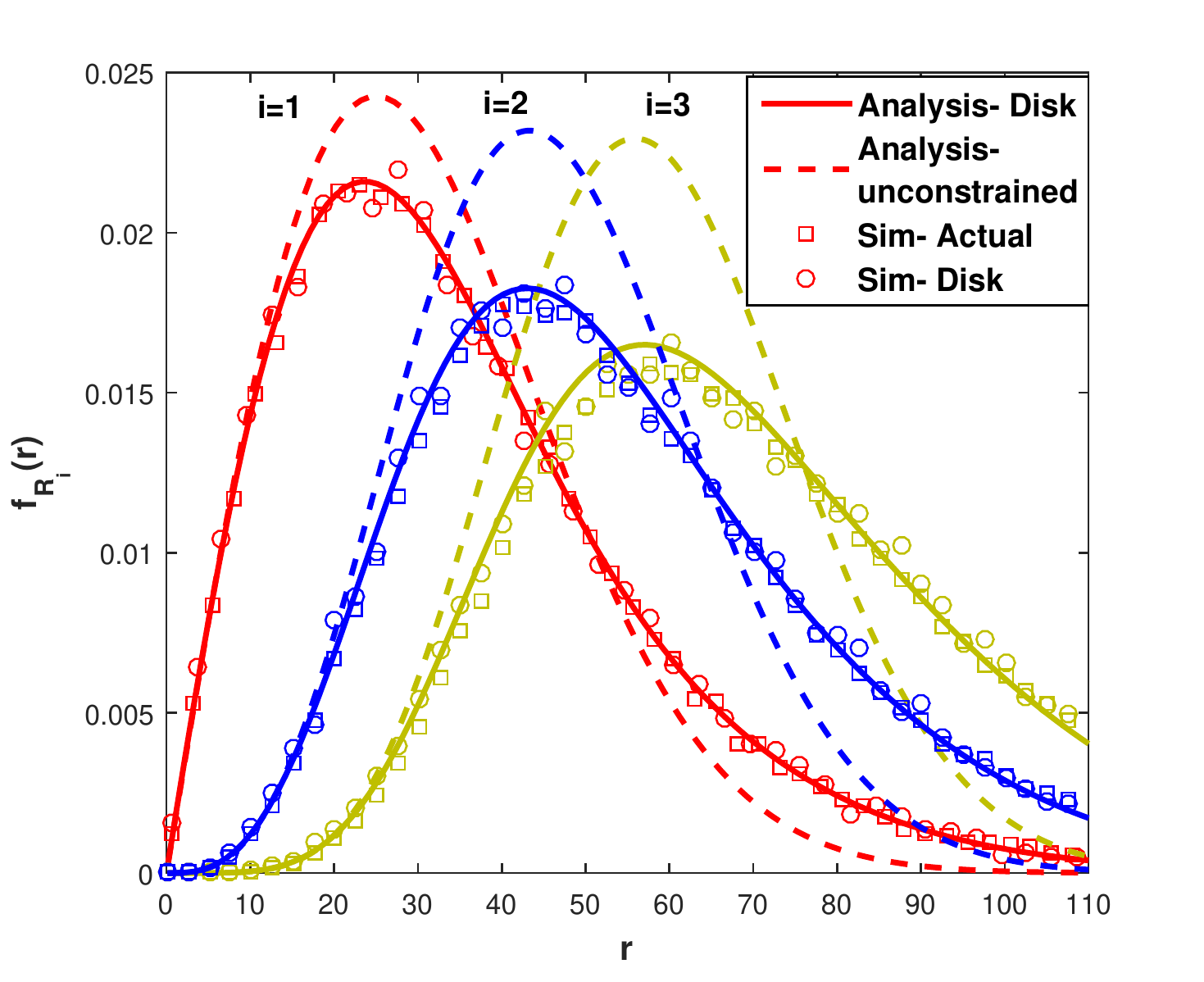}\caption{Distribution of the distance to the $i$th nearest D2D helper from
the tagged user within the Voronoi cell, where $\lambda_{m}=20/\pi500^{2},$
and $\lambda_{d}=200/\pi500^{2}$.\label{fig:Dist}}
\end{figure}

We first validate the distribution of distance to the $i$th nearest
D2D helper derived in Theorem \ref{prop:dist} for various values
of $i$. For the simulations, we ignore the realizations in which
the number of D2D helpers is less than $i$ in the typical cell. In
case of the disk approximation, the realizations in which the typical
user lies outside $B_{max}$, or there are less than $i$ D2D helpers
inside $B_{max}$, are all ignored. Fig. \ref{fig:Dist} shows that
the disk approximation is very accurate, while the unconstrained nearest
neighbor distribution in (\ref{eq:f2r}) does not encapsulate the
behavior of the distance distribution and the deviations from the
actual distribution become large as the value of $i$ increases.

\begin{figure}
\centering{}\includegraphics[scale=0.65]{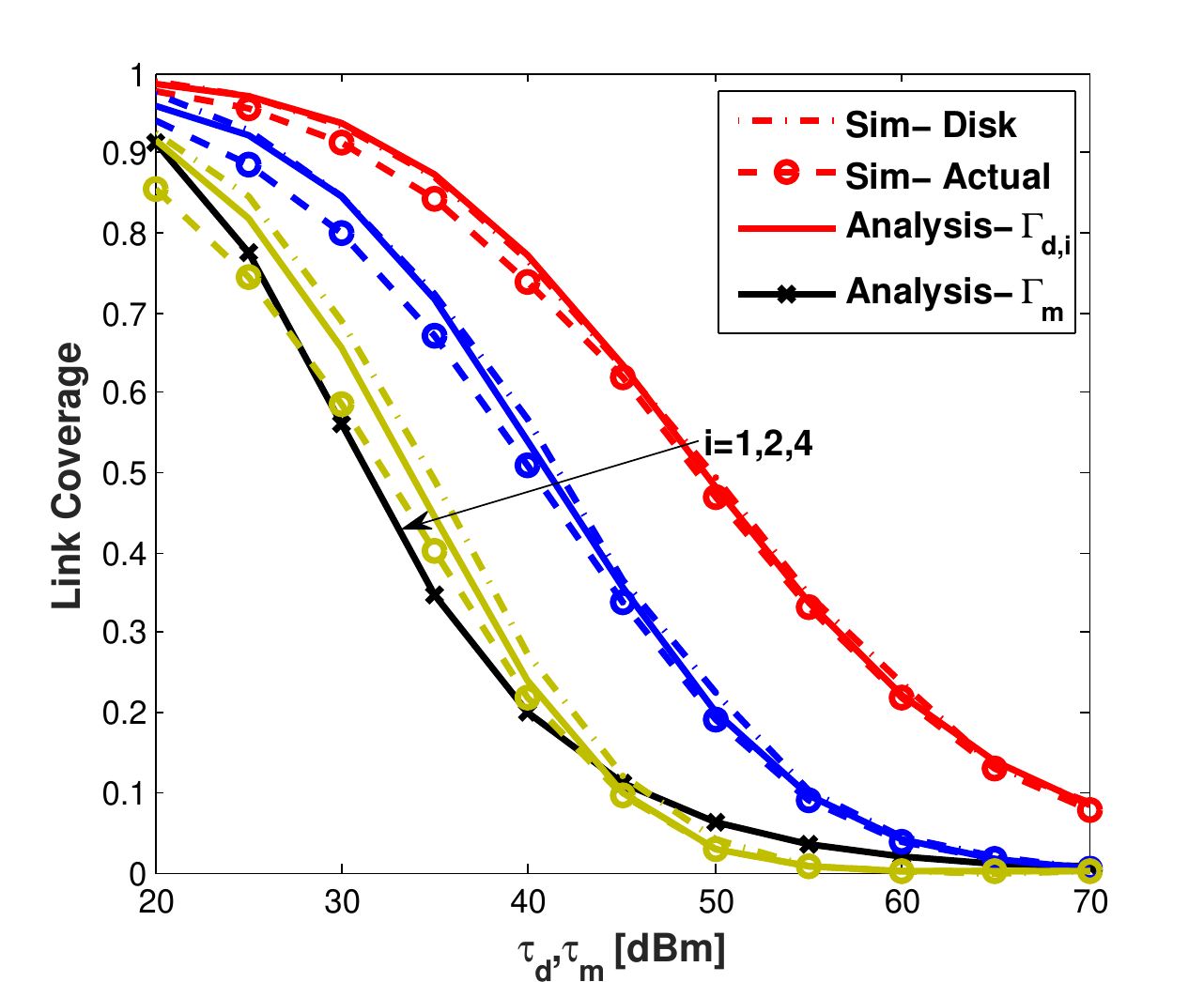}\caption{Probability of coverage when a typical user is served by the MBS or
the $i$th nearest D2D helper. \label{fig:covD}}
\end{figure}

Fig. \ref{fig:covD} validates our analysis of the probability of
coverage $\Gamma_{d,i}$ when the typical user is being served by
the $i$th nearest D2D helper ((\ref{eq:covD2D}) in Theorem \ref{prop:covD}).
We see that the disk approximation holds fairly accurately for all
values of $i$. The slight deviation of the analysis using the disk
approximation from the simulations using disk approximation is because
of the equi-dense HPPP approximation for the D2D interferers. As expected,
we see a decrease in $\Gamma_{d,i}$ with the increases in $i$ for
a fixed SINR threshold. This is because, as $i$ increases, the distance
between the transmitting D2D helper and the typical user increases,
thereby aggravating the path loss. For comparison, we also plot the
cellular coverage $\Gamma_{m}$ given in (\ref{eq:covC}). For a given
SINR threshold, we see that small values of $i$ result in a much
better coverage for a D2D link compared to the cellular link. 

\begin{figure}
\centering{}\includegraphics[scale=0.65]{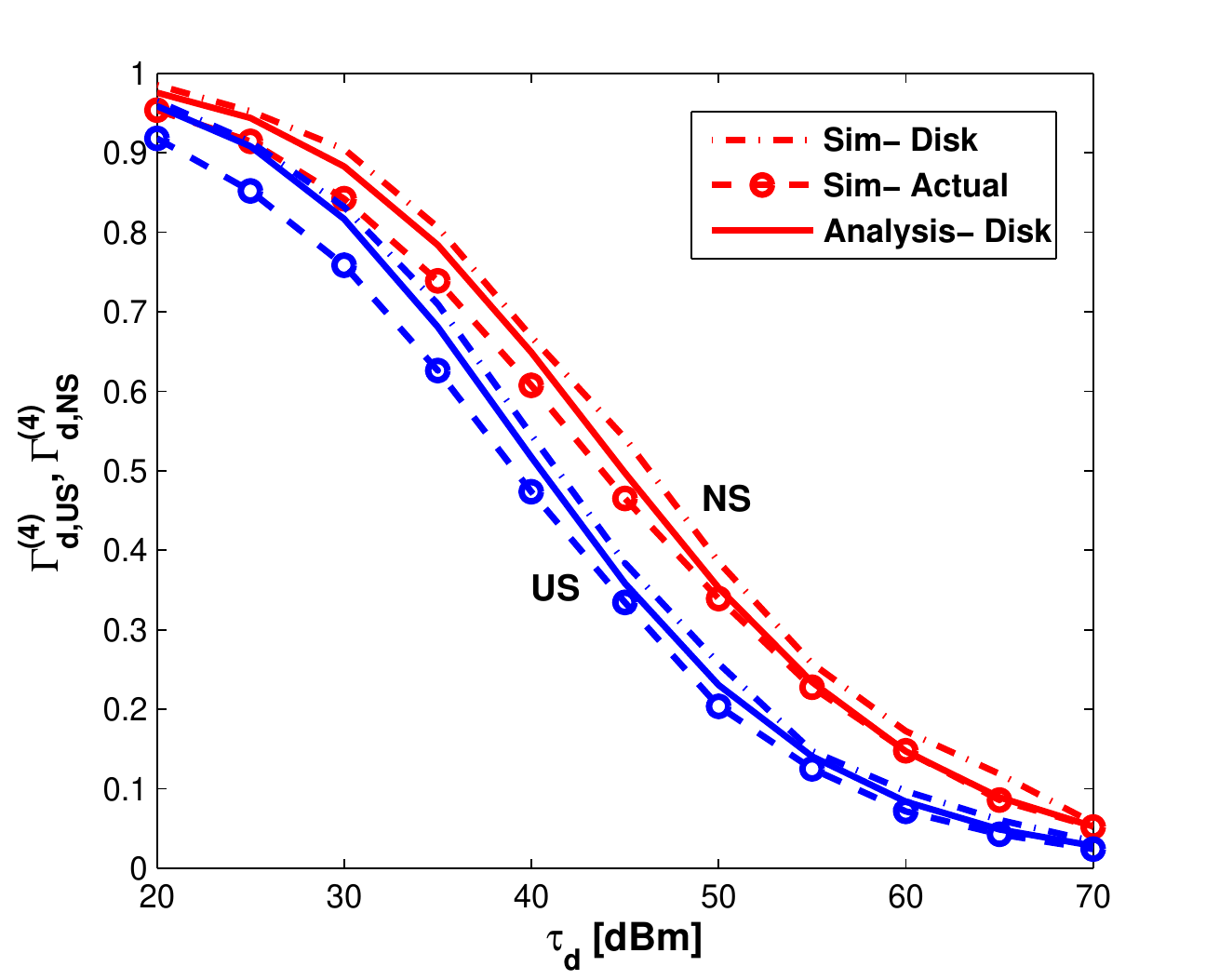}\caption{\label{fig:Overall-D2D-coverage}Coverage probability in D2D mode
under the NS and US schemes. }
\end{figure}

Fig. \ref{fig:Overall-D2D-coverage} illustrates the probability of
being in coverage in D2D mode. We also validate the D2D coverage probability
$\Gamma_{d,\varPi}^{(k)}(c),\varPi=\{NS,US\}$ derived in Theorem
\ref{prop:covD} (\ref{eq:covDk}). For each simulation trial, maximum
$k$ closest D2D helpers are first checked for content availability.
The content availability ($c=1$ in this case) is a Bernoulli event
with probability $h_{d}(c)$. Out of the successful D2D helpers (if
there are any), the helper is selected either uniformly (US scheme)
or closest to the origin (NS scheme). We see that the D2D coverage
probability for the NS scheme outperforms the US scheme. This is because
in the US scheme, a D2D helper is uniformly selected out of the maximum
$k$ closest helpers, while the closest helper is given preference
in the NS scheme. 
\begin{figure}[b]
\subfloat{\rule[0.5ex]{0.75\paperwidth}{0.9pt}}
\begin{equation}
\Gamma_{US}^{(k)}(c)\approx\left[1-\rho c^{-\zeta}\right]^{C_{d}}\Gamma_{m}+\frac{1}{k}\left(1-\left[1-\rho c^{-\zeta}\right]^{C_{d}}\right)\sum_{i=1}^{k}\Gamma_{d,i}\label{eq:CovUSsimp}
\end{equation}
\begin{equation}
\Gamma_{NS}^{(k)}(c)\approx\left[1-\rho c^{-\zeta}\right]^{kC_{d}}\Gamma_{m}+\left(1-\left[1-\rho c^{-\zeta}\right]^{C_{d}}\right)\sum_{i=1}^{k}\left[1-\rho c^{-\zeta}\right]^{(i-1)C_{d}}\Gamma_{d,i}\label{eq:CovNSsimp}
\end{equation}
\end{figure}

\begin{figure}
\centering{}\includegraphics[scale=0.65]{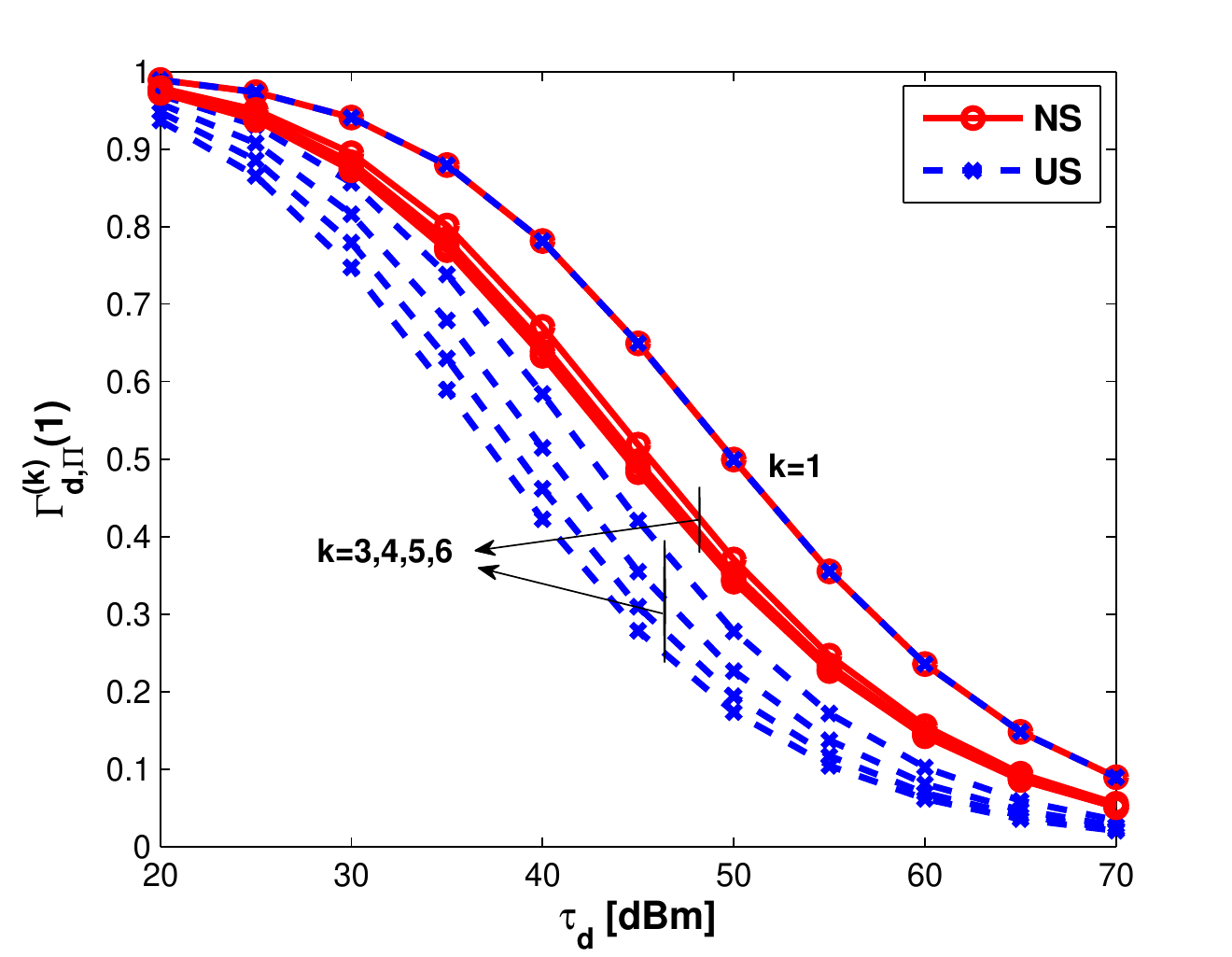}\caption{\label{fig:D2D-coverage-for}D2D coverage for various values of $k$
(increasing to the left). }
\end{figure}

The behavior of the D2D coverage probability $\Gamma_{d,\varPi}^{(k)}(c),\varPi=\{NS,US\}$
is further investigated when the value of $k$ is changed. We can
see from Fig. \ref{fig:D2D-coverage-for} that the increase in $k$
adversely affects $\Gamma_{d,US}^{(k)}(c)$. However, the effect on
$\Gamma_{d,NS}^{(k)}(c)$ is much less pronounced. 
\begin{figure}
\centering{}\includegraphics[scale=0.6]{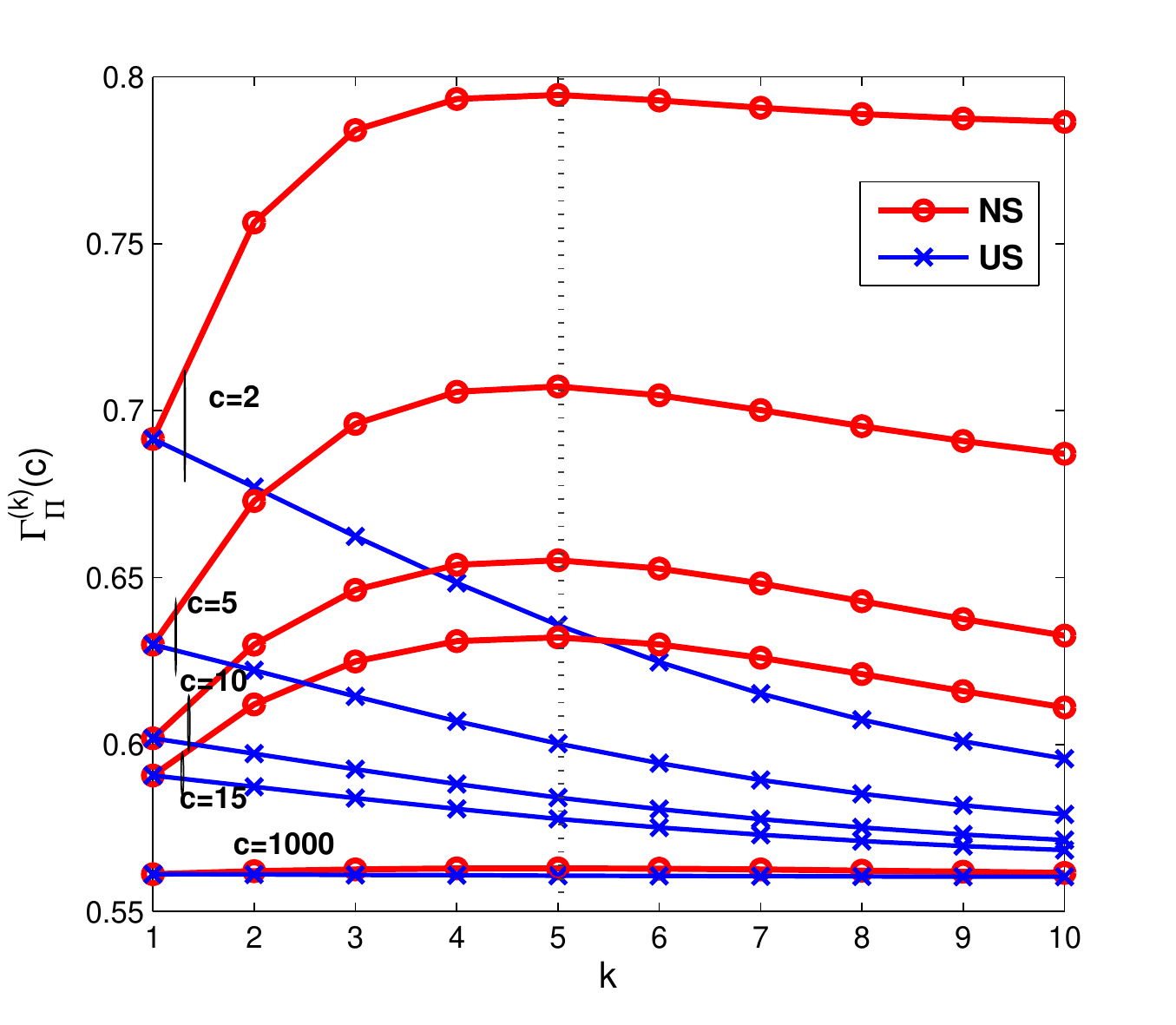}\caption{\label{fig:OverallCov}Effect of increasing $k$ on the overall coverage
probability for various content requests. }
\end{figure}
\begin{figure}
\begin{centering}
\includegraphics[scale=0.6]{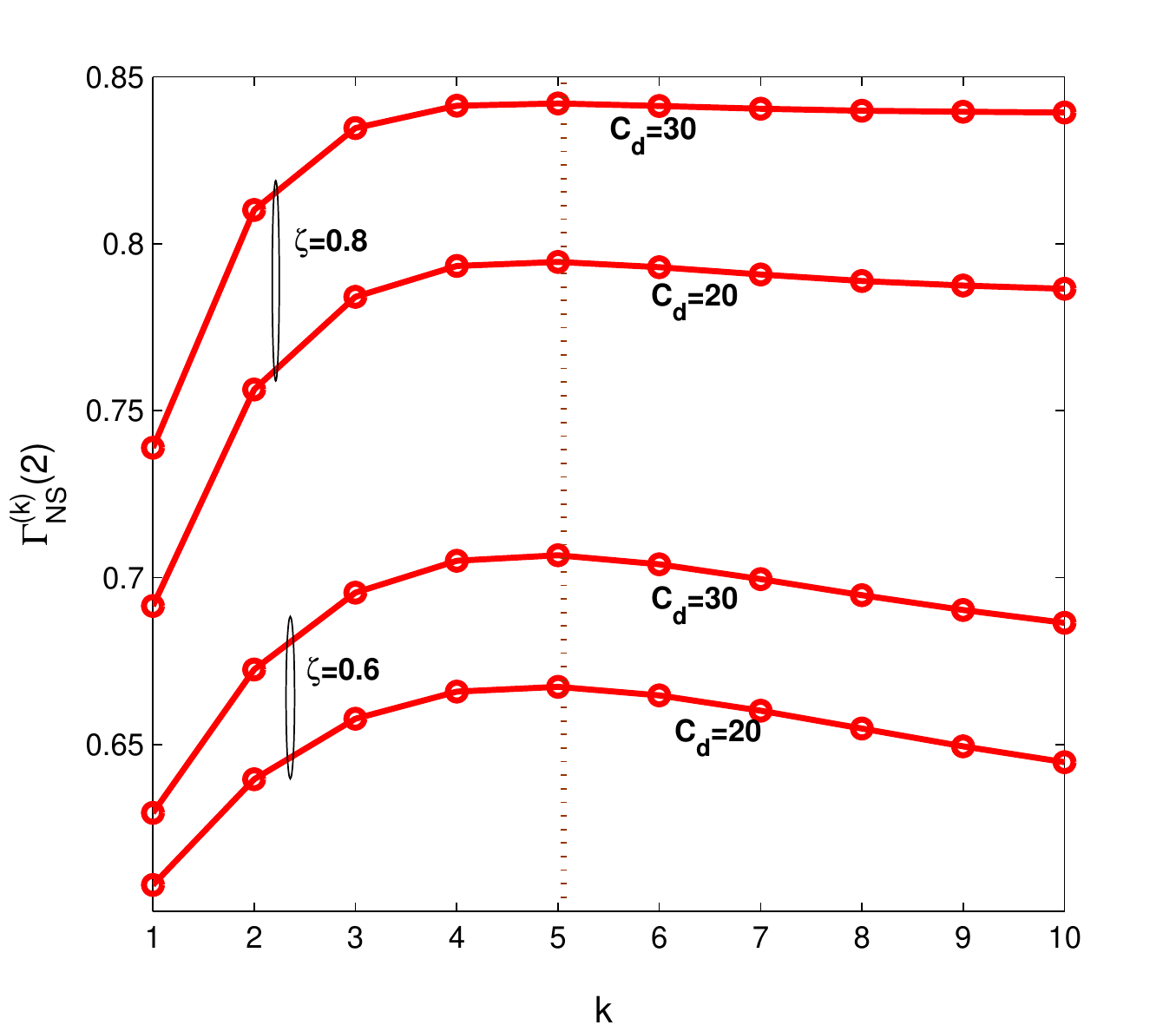}
\par\end{centering}
\caption{Effect of D2D caching parameters on $k_{NS}^{*}$.\label{fig:kOptNS}}
\end{figure}
\begin{figure}
\begin{centering}
\includegraphics[scale=0.6]{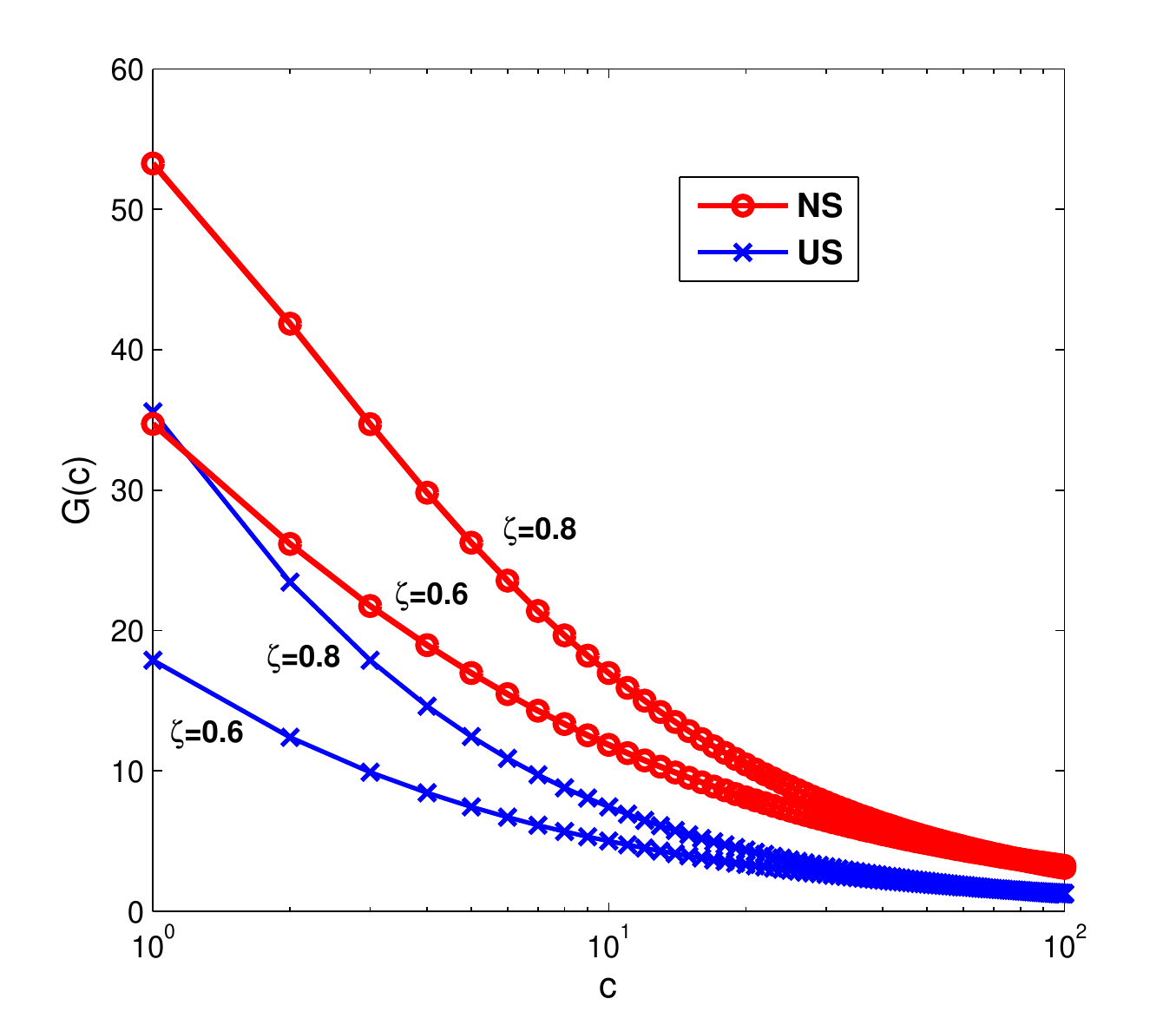}
\par\end{centering}
\caption{Percentage maximum gain in the overall coverage with coordinated D2D
under NS and US schemes.\label{fig:Percentage-gain-in}}
\end{figure}

\subsection{Performance Evaluation}

\begin{figure}[b]
\subfloat{\rule[0.5ex]{0.75\paperwidth}{0.9pt}}
\begin{equation}
T_{US}^{(k)}(c)\approx\left[1-\rho c^{-\zeta}\right]^{C_{d}}W_{m}\gamma\left(\eta_{u_{m,US}}\right)\hat{R}_{m}(c)+\frac{W_{d}}{k}\gamma\left(\eta_{u_{d,US}}\right)\left(1-\left[1-\rho c^{-\zeta}\right]^{C_{d}}\right)\sum_{i=1}^{k}R_{d,i}\label{eq:tpsimpUS}
\end{equation}
\begin{eqnarray}
T_{NS}^{(k)}(c) & \approx & \left[1-\rho c^{-\zeta}\right]^{kC_{d}}W_{m}\gamma\left(\eta_{u_{m,NS}}\right)\hat{R}_{m}(c)\nonumber \\
 &  & +W_{d}\gamma\left(\eta_{u_{d,NS}}\right)\left(1-\left[1-\rho c^{-\zeta}\right]^{C_{d}}\right)\sum_{i=1}^{k}\left[1-\rho c^{-\zeta}\right]^{(i-1)C_{d}}R_{d,i}\label{eq:tpsimpNS}
\end{eqnarray}
\end{figure}
We now study the performance metrics defined in Sec. \ref{sec:Link-Spectral-Efficiency}
with respect to the two key parameters, namely, the number of candidate
neighboring D2D helpers $k$ and the requested content $c$. We first
consider the overall coverage probability $\Gamma_{\varPi}^{(k)}(c),\varPi=\{NS,US\}$
given in (\ref{eq:OverallCov}). The simplified expressions for $\Gamma_{\varPi}^{(k)}(c)$
are presented in (\ref{eq:CovUSsimp}) and (\ref{eq:CovNSsimp}) using
the upper bounds for $p_{\varPi}^{(k)}(i,c)$ from Corollary \ref{cor:hitSimple}.
In Fig. \ref{fig:OverallCov}, the overall coverage probability for
the US scheme $\Gamma_{US}^{(k)}(c)$ is seen to monotonically decrease
with the increase in $k$, starting with a maximum value at $k_{US}^{*}=1$.
This is because, from Fig. \ref{fig:covD} we see that the D2D link
coverage $\Gamma_{d,i}$ decreases as $i$ increases and even gets
worse compared to the cellular link coverage $\Gamma_{m}$. As $k$
increases in (\ref{eq:CovUSsimp}), the contribution of $\Gamma_{d,1}$
decreases in the second term corresponding to the coverage in D2D
mode. Intuitively, this means that the MBS does not make an intelligent
decision in the selection of the D2D helper and may select a helper
farther from the requesting user for D2D communication. As the first
term in (\ref{eq:CovUSsimp}) is independent of $k$, and $\Gamma_{d,i}$
is a decreasing function in $i$, (\ref{eq:CovUSsimp}) is always
maximized when $k_{US}^{*}=1$. On the contrary, the plots for $\Gamma_{NS}^{(k)}(c)$
reveal an interesting trade off in the selection of $k$ to maximize
the overall coverage. We see that for a given SINR threshold (and
also $\tau_{d}=\tau_{m}$), there exists an optimal value of $k=k_{NS}^{*}$,
which maximizes $\Gamma_{NS}^{(k)}(c)$ (shown by the dotted line).
The explanation for this phenomenon is given as follows. We see that
as $k$ is increased, the probability of cellular mode is decreased
as the term $\left[1-\rho c^{-\zeta}\right]^{kC_{d}}$ premultiplied
with $\Gamma_{m}$ in (\ref{eq:CovNSsimp}) decreases. Initially,
$\Gamma_{NS}^{(k)}(c)$ rises with the increase in $k$ and attains
a maximum value at $k_{NS}^{*}$, but with further increase in $k$,
the coverage $\Gamma_{d,i}$ with the activated D2D links is no longer
better than the cellular coverage as already seen from Fig. \ref{fig:covD},
hence, $\Gamma_{NS}^{(k)}(c)$ begins to drop. We observe that as
the requested content $c$ becomes less popular, both $p_{NS}^{(k)}(c)$
and $p_{US}^{(k)}(c)$ drop, and for the least popular requested content
$\Gamma_{\varPi}^{(k)}(c)\rightarrow\Gamma_{m}$ as $h_{d}(c)\rightarrow0$
implying that the requesting user can only be served in cellular mode.
It can also be seen that varying $c$ does not affect the optimal
value $k_{NS}^{*}$. In Fig. \ref{fig:kOptNS}, we also observe the
effect of the other crucial D2D caching parameters, $\zeta$ and $C_{d}$
on $\Gamma_{NS}^{(k)}(c)$. We see that $k_{NS}^{*}$ is also resilient
to the changes in $\zeta$ and $C_{d}$, but the maximum value of
the overall coverage $\Gamma_{NS}^{(k)}(c)$ increases with the increase
in $\zeta$ and $C_{d}$. This is due to the fact that popular content
$c$ or higher values of $\zeta$ and $C_{d}$ all translate into
a higher probability of being served by the $i$th nearest D2D helper
as $p_{\varPi}^{(k)}(i,c)$ increases $\forall i=\{1,..,k\}$, but
these parameters do not affect the link quality ($\Gamma_{m}$ and
$\Gamma_{d,i}$). Therefore, $k_{NS}^{*}$ is independent of the caching
parameters. 

To better visualize the improvement in coverage compared to the conventional
cellular network scenario, we compute the gain in the overall coverage
probability. It is the percentage difference between the maximum attainable
overall coverage probability under the NS and US schemes $\Gamma_{\varPi}^{(k_{\varPi}^{*})}(c),\varPi=\{NS,US\}$
and the conventional cellular coverage probability. It is given as

\[
G(c)=\frac{\Gamma_{\varPi}^{(k_{\varPi}^{*})}(c)-\Gamma_{m}}{\Gamma_{m}}\times100\%.
\]
Figure \ref{fig:Percentage-gain-in} shows that for popular content
requests and skewed popularity distribution, more than 50\% better
coverage can be obtained with the NS scheme compared to the conventional
scenario. The US scheme does not perform as good as the NS scheme,
but it yields sufficient gains (35\% at best under the given network
setting). 
\begin{figure}
\centering{}\includegraphics[scale=0.7]{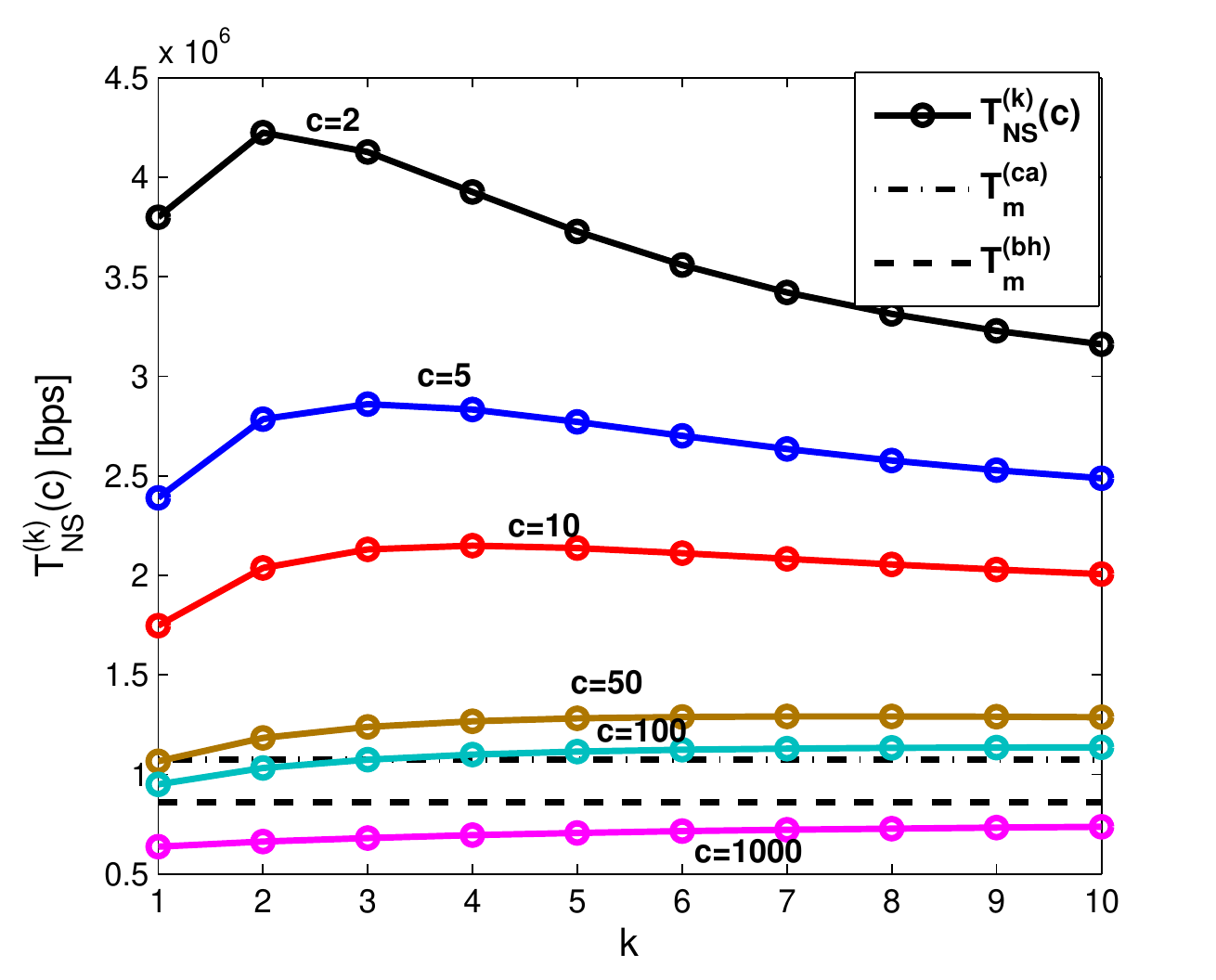}\caption{Effect of increasing $k$ on the average rate experienced by an arbitrary
user for various content requests.\label{fig:AvgRateC}}
\end{figure}

We will now focus on the analysis of average rate experienced by the
user in the US and NS schemes to gain some more useful insights on
the design of cellular networks enabled with coordinated D2D communication.
The simplified expressions for $T_{US}^{(k)}(c)$ and $T_{NS}^{(k)}(c)$
using Corollary \ref{cor:hitSimple} are given in (\ref{eq:tpsimpUS})
and (\ref{eq:tpsimpNS}) respectively. For the average rate with the
US scheme, we see from (\ref{eq:tpsimpUS}) that $T_{US}^{(k)}(c)$
exhibits the same behavior as $\Gamma_{US}^{(k)}(c)$ the increase
in $k$ reduces the average rate in D2D mode because D2D helpers located
farther from the requesting user are selected with the probability
equal to the nearest helpers. Hence, $\Gamma_{US}^{(k)}(c)$ is only
maximized when $k=k_{US}^{*}=1$. 

To visualize when maximum gains can be harnessed with the NS scheme,
we plot the $\Gamma_{NS}^{(k)}(c)$ in Fig. \ref{fig:AvgRateC} and
compare it with the rates experienced by the user in a cellular only
network. The rate experienced by the user in a cellular network where
the MBS is equipped with caching capability is given as
\begin{equation}
T_{m}^{(ca)}=(W_{m}+W_{d})\gamma\left(\eta_{u}\right)R_{m}.
\end{equation}
 Notice that all of the bandwidth $W_{d}+W_{c}$ is used for cellular
communication and $\eta_{u}=\lambda_{u}/\lambda_{m}$ as all the requesting
users in the cell share the cellular link capacity. When there is
no caching at the MBS, the rate is reduced by a fraction $\beta$
due to the delay introduced by the backhaul communication and is given
as 
\begin{equation}
T_{m}^{(bh)}=\beta T_{m}^{(ca)}.
\end{equation}
The following conclusions can be drawn from Fig. \ref{fig:AvgRateC}.
Coordinated D2D communication with content centric mode selection
greatly enhances data rates compared to cellular only scenarios when
popular contents are requested. This is because we know that for small
$c$, $h_{d}(c)$ (and in turn $p_{NS}^{(k)}(i,c)$) increases implying
that more users can be served by their nearest D2D helper in D2D mode.
For least popular contents, the rate experienced by the requesting
user is severely degraded and is even lower than the cellular only
scenario. This is because, $h_{d}(c)\rightarrow0$ and the user is
pushed to communicate in cellular band $W_{m}$, which is less than
the total bandwidth $W_{m}+W_{d}$ of the cellular only scenario.
Yet again, there exists a trade off in the selection of the number
of neighboring D2D helpers $k$ which maximizes the average throughput
experienced by an arbitrary user. The existence of the optimal value
of $k=k_{NS}^{*}$ for $T_{NS}^{(k)}(c)$ follows the same reasoning
as $\Gamma_{NS}^{(k)}(c)$, as $R_{d,i}$ follows the same trend as
$\Gamma_{d,i}$ and is decreasing in $i.$ However, Unlike the $\Gamma_{NS}^{(k)}(c)$,
this value of $k=k_{NS}^{*}$ does vary with the changes in $c$ and
increases as $c$ increases. This is because, as we have from Proposition
\ref{prop:Rd} 
\begin{eqnarray*}
\eta_{u_{d,NS}} & = & \frac{\lambda_{u}}{\lambda_{m}}\left[\rho\sum_{c=1}^{L}c^{-\zeta}p_{d,NS}^{(k)}(c)\right]\\
 & = & \frac{\lambda_{u}}{\lambda_{m}}\left[1-\rho\sum_{c=1}^{L}c^{-\zeta}[1-\rho c^{-\zeta}]^{kC_{d}}\right],
\end{eqnarray*}
where $\gamma\left(\eta_{u_{d,NS}}\right)=\eta_{u_{d,\varPi}}^{-1}\left(1-\textnormal{exp}\left(-\eta_{u_{d,\varPi}}\right)\right)$
is a decreasing function in $k$. This means that more users are offloaded
from cellular to D2D mode as $k$ increases and the available bandwidth
\textbf{$W_{d}\gamma\left(\eta_{u_{d,NS}}\right)$ }for a user in
D2D mode decreases. When $k\rightarrow\infty,$ $\gamma\left(\eta_{u_{d,NS}}\right)\rightarrow\gamma\left(\eta_{u}\right)$
implying that all the requesting users are served in D2D mode. We
know that as $c$ increases, then the term $1-\left[1-\rho c^{-\zeta}\right]^{C_{d}}$
decreases in (\ref{eq:tpsimpNS}) and the D2D rate decreases. To effectively
utilize the D2D bandwidth $W_{d}$, more users need to be activated
in D2D mode and hence, $k_{NS}^{*}$ increases. 

\section{Conclusion\label{sec:Conclusion}}

In this paper, we presented a novel framework for the analysis of
cache-enabled cellular networks with coordinated D2D communication.
The arbitrary user requesting a particular content is offloaded to
communicate with one of its $k$ neighboring D2D helpers within the
cell based on the content availability and helper selection schemes.
We derived the distribution of the distance between the user and its
$i$th nearest D2D helper within the cell using disk cell approximation,
which is shown to be fairly accurate. We obtained the probabilities
for being served in cellular and D2D modes and the coverage and data
rates experienced by the user in both these modes. With the help of
our analysis, we showed that the information-centric offloading with
coordinated D2D results in high performance gain. However, to maximize
the performance, the number of candidate D2D helpers $k$ has to be
carefully tuned.

\appendices{}

\section{Proof of Theorem\label{sec:Proof1} \ref{prop:dist}}

\begin{figure}
\begin{centering}
\subfloat[]{\centering{}\includegraphics[scale=0.6]{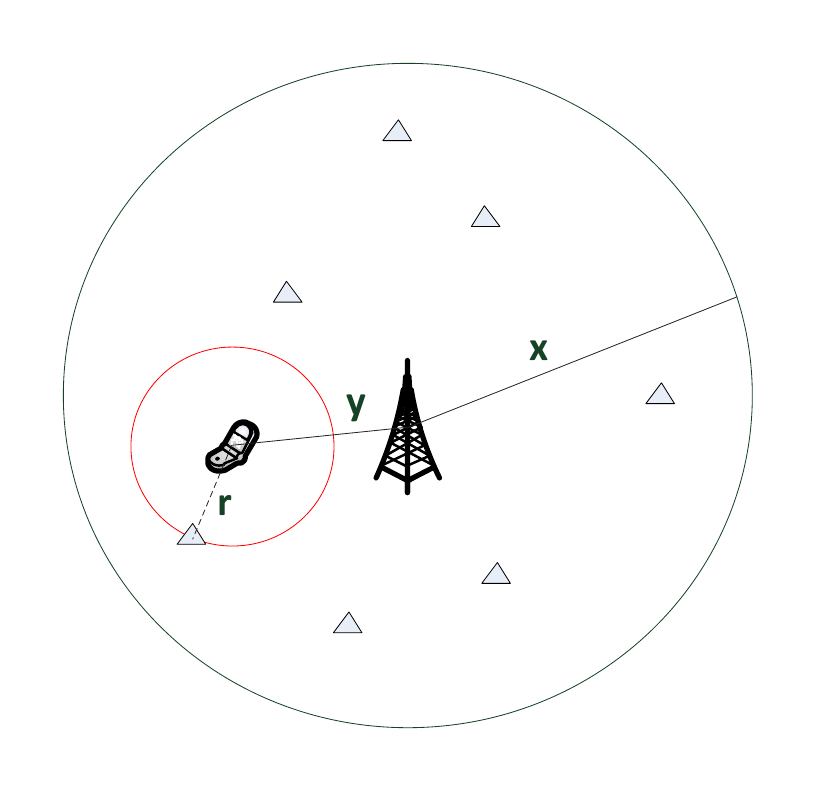}}\enskip{}\subfloat[]{\centering{}\includegraphics[scale=0.6]{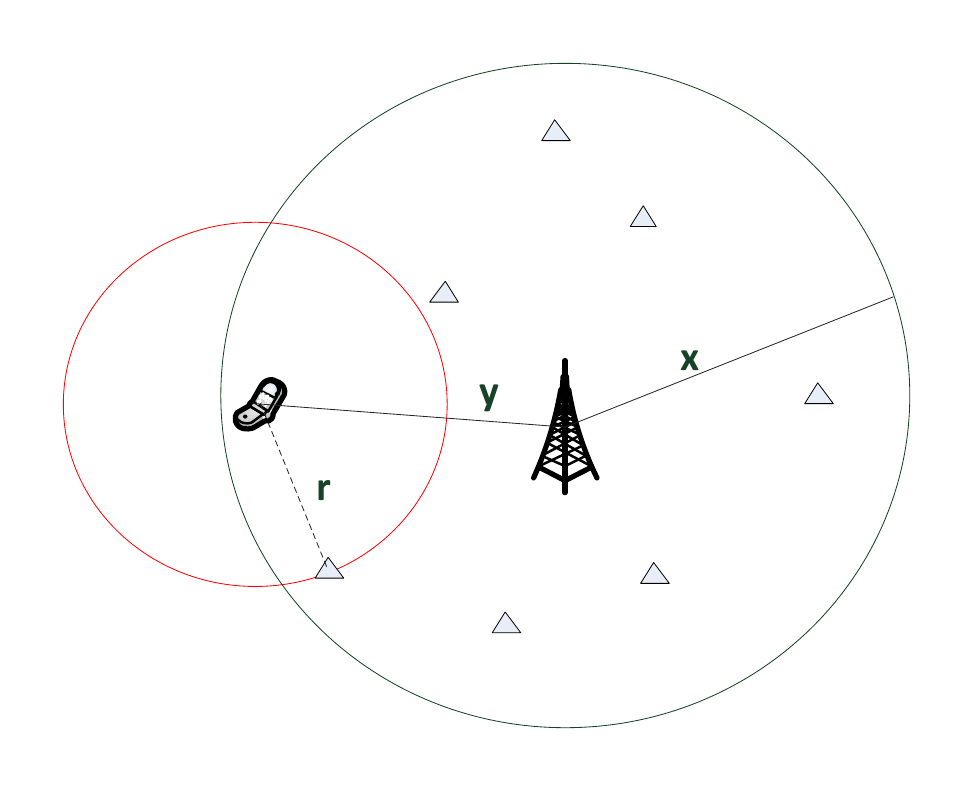}}
\par\end{centering}
\caption{Distance to the nearest D2D neighbor within the circular Voronoi cell:
(a) when $b(o,r)$ is inside $B_{max}$, (b) when $b(o,r)$ partly
overlaps $B_{max}$. \label{fig:distnearest} }
\end{figure}

The probability that the distance between the requesting user and
the $i$th nearest D2D helper within the cell is at least $r$ is
the probability that there are exactly $i-1$ helpers inside the region
$\mathcal{A}$. It can be expressed as

\begin{equation}
1-F_{R_{i}|X=x,X>D,N_{d}\geq i}(r)=\frac{\left(\lambda_{d}\mathcal{A}\right)^{i-1}}{(i-1)!}\,\textnormal{exp}\left(-\lambda_{d}\mathcal{A}\right),\label{eq:cdflens}
\end{equation}
where $\mathcal{A}$ is the area of intersection between $B_{m}$
and $b(o,r).$ As shown in Fig. \ref{fig:distnearest}, this area
can be divided into two regimes given as follows.
\begin{itemize}
\item Regime 1 - When $b(o,r)$ partly overlaps $B_{m}$, i.e. $x-y<r<x+y$.
The overlapping area $\mathcal{A}$ in this case can be written as
\cite{weisstein2003circle} \\
\begin{eqnarray}
\nabla(r,y,x) & = & r^{2}\arccos\left(\frac{\omega_{1}}{2y\,r}\right)+x^{2}\arccos\left(\frac{\omega_{2}}{2y\,x}\right)-\frac{1}{2}\sqrt{4y^{2}x^{2}-\omega_{2}^{2}},\label{eq:lens}
\end{eqnarray}
where $\omega_{1}=r^{2}+y^{2}-x^{2}$ and $\omega_{2}=x^{2}+y-r^{2}$. 
\item Regime 2 - When $b(o,r)$ lies inside $B_{m}$ i.e. $0<r<x-y$. The
overlapping area in this case is straightforward and is given as $\mathcal{A}=\pi r^{2}.$
\end{itemize}
Differentiating $F_{R_{i}|X=x,X>D,N_{d}\geq i}(r)$ in (\ref{eq:cdflens})
with respect to $r$ gives (\ref{eq:f1r}) and (\ref{eq:f2r}) for
regimes 1 and 2 respectively. The unconditional distance distribution
$f_{R_{i}}(r)$ in (\ref{eq:fr}) is obtained by averaging over $X$
and $Y$, where $X>Y$ and $N_{d}\geq i$.

\section{Proof of Proposition\label{sec:Proof2} \ref{prop:covD}}

The probability of coverage for the typical user served by the $i$th
nearest D2D helper can be expressed as
\begin{eqnarray}
\varGamma_{d,i} & = & \mathbb{P}\left\{ \frac{h_{i}\,r^{-\alpha}}{\sigma^{2}/P_{d}+I_{d}}>\tau_{d}\right\} \label{eq:covD}
\end{eqnarray}
where $I_{d}=\sum_{z_{j}\in\Phi_{d}^{int}}g_{j}\,z_{j}^{-\alpha}$
is the inter-cell interference from other active D2D helpers, which
is the sum of powers from the active D2D helpers constituting $\Phi_{d}^{int}$.
Here, $z_{j}$ is the distance of the the interfering D2D helper from
the typical user and $g_{j}$ is the channel power for the interfering
link $j$. Because of the exponentially distributed channel power
$h_{i}$ in (\ref{eq:covD}), we get
\begin{equation}
\varGamma_{d,i}=\mathbb{E}_{R_{i}}\left[\exp\left(-s_{d}\sigma^{2}/P_{d}\right)\mathcal{L}_{I_{d}}\left(s_{d}\right)\right],\label{eq:Covdi}
\end{equation}
where $s_{d}=\tau_{d}r^{\alpha}$ and $\mathcal{L}_{I_{d}}\left(s_{d}\right)=\mathbb{E}_{I_{d}}\left[\exp\left(-s_{d}I_{d}\right)\right]$
is the Laplace transform of D2D interference. Because only one D2D
helper can be active at one channel in a given macrocell, we employ
a key assumption that $\Phi_{d}^{int}$ is a HPPP with intensity $\tilde{\lambda}_{m}=p_{int}\times\lambda_{m}$
\footnote{The equi-dense HPPP assumptions ignores the correlations due to the
position of helpers inside a cell, but is more tractable. \cite{elsawy2014analytical}}. Here, $p_{int}=1-\left(1+3.5^{-1}\eta_{d}\right)^{-3.5}$ is the
probability that at least one interfering D2D helper is present in
a cell. $\mathcal{L}_{I_{d}}\left(s_{d}\right)$ can then be presented
as 
\begin{eqnarray}
\mathcal{L}_{I_{d}}\left(s_{d}\right) & = & \mathbb{E}\left[\text{exp}\left(-s_{d}\sum_{\textbf{z}_{j}\in\Theta_{d}}g_{j}\,z_{j}^{-\alpha}\right)\right]\nonumber \\
 & \overset{(a)}{=} & \mathbb{E}_{Q}\biggl[\exp\biggl(-2\pi\tilde{\lambda}_{m}\intop_{q}^{\infty}\frac{\nu}{1+s_{d}^{-1}\nu^{\alpha}}\,d\nu\biggr)\biggr]\label{eq:laplaceD2D}
\end{eqnarray}
where $(a)$ follows from the generating functional of a HPPP and
the exponential distribution of the channel power $g$. The lower
limit of the integral in (\ref{eq:laplaceD2D}) represents the guard
zone. Notice that the lower limit $q$ in this case is governed by
the nearest active D2D interferer, where $f_{Q}(q)=2\pi\tilde{\lambda}_{m}q\,\text{exp}(-\tilde{\lambda}_{m}\pi q^{2})$
because of the equi-dense HPPP approximation. As $\tilde{\lambda}_{M}$
is quite small, we can apply Jensen's inequality to achieve a tight
bound for (\ref{eq:laplaceD2D}) 

\begin{equation}
\mathcal{L}_{I_{d}}\left(s_{d}\right)\thickapprox\exp\left(-2\pi\tilde{\lambda}_{m}\mathbb{E}_{Q}\left[\intop_{q}^{\infty}\frac{\nu}{1+s_{d}^{-1}\nu^{\alpha}}\,d\nu\right]\right).\label{eq:lapDJen}
\end{equation}
Substituting (\ref{eq:lapDJen}) into (\ref{eq:Covdi}) gives (\ref{eq:covD2D}).

The overall D2D coverage probability for NS and US schemes and a particular
content request in (\ref{eq:covDk}) is obtained by taking expectation
over $i$ and by conditioning over the probability of D2D mode.

\bibliographystyle{IEEEtran}
\bibliography{D2Dref}

% Generated by IEEEtran.bst, version: 1.14 (2015/08/26)
\begin{thebibliography}{10}
\providecommand{\url}[1]{#1}
\csname url@samestyle\endcsname
\providecommand{\newblock}{\relax}
\providecommand{\bibinfo}[2]{#2}
\providecommand{\BIBentrySTDinterwordspacing}{\spaceskip=0pt\relax}
\providecommand{\BIBentryALTinterwordstretchfactor}{4}
\providecommand{\BIBentryALTinterwordspacing}{\spaceskip=\fontdimen2\font plus
\BIBentryALTinterwordstretchfactor\fontdimen3\font minus
  \fontdimen4\font\relax}
\providecommand{\BIBforeignlanguage}[2]{{%
\expandafter\ifx\csname l@#1\endcsname\relax
\typeout{** WARNING: IEEEtran.bst: No hyphenation pattern has been}%
\typeout{** loaded for the language `#1'. Using the pattern for}%
\typeout{** the default language instead.}%
\else
\language=\csname l@#1\endcsname
\fi
#2}}
\providecommand{\BIBdecl}{\relax}
\BIBdecl

\bibitem{bastug2014living}
E.~Bastug, M.~Bennis, and M.~Debbah, ``Living on the {E}dge: {T}he {R}ole of
  {P}roactive {C}aching in 5{G} {W}ireless {N}etworks,'' \emph{IEEE
  Communications Magazine}, vol.~52, no.~8, pp. 82--89, 2014.

\bibitem{woo2013comparison}
S.~Woo, E.~Jeong, S.~Park, J.~Lee, S.~Ihm, and K.~Park, ``{C}omparison of
  {C}aching {S}trategies in {M}odern {C}ellular {B}ackhaul {N}etworks,'' in
  \emph{Proceeding of the 11th annual international conference on Mobile
  systems, applications, and services}.\hskip 1em plus 0.5em minus 0.4em\relax
  ACM, 2013, pp. 319--332.

\bibitem{lin2014overview}
X.~Lin, J.~Andrews, A.~Ghosh, and R.~Ratasuk, ``An {O}verview of 3{GPP}
  {D}evice-to-{D}evice {P}roximity {S}ervices,'' \emph{IEEE Communications
  Magazine}, vol.~52, no.~4, pp. 40--48, 2014.

\bibitem{malandrino2014toward}
F.~Malandrino, C.~Casetti, and C.-F. Chiasserini, ``{T}oward {D2D}-{E}nhanced
  {H}eterogeneous {N}etworks,'' \emph{IEEE Communications Magazine}, vol.~52,
  no.~11, pp. 94--100, 2014.

\bibitem{fodor2012design}
G.~Fodor, E.~Dahlman, G.~Mildh, S.~Parkvall, N.~Reider, G.~Mikl{\'o}s, and
  Z.~Tur{\'a}nyi, ``Design {A}spects of {N}etwork {A}ssisted
  {D}evice-to-{D}evice {C}ommunications,'' \emph{IEEE Communications Magazine},
  vol.~50, no.~3, pp. 170--177, 2012.

\bibitem{asadi2014survey}
A.~Asadi, Q.~Wang, and V.~Mancuso, ``A {S}urvey on {D}evice-to-{D}evice
  {C}ommunication {I}n {C}ellular {N}etworks,'' \emph{IEEE Communications
  Surveys \& Tutorials}, vol.~16, no.~4, pp. 1801--1819, 2014.

\bibitem{haenggi2012stochastic}
M.~Haenggi, \emph{Stochastic {G}eometry for {W}ireless {N}etworks}.\hskip 1em
  plus 0.5em minus 0.4em\relax Cambridge University Press, 2012.

\bibitem{ji2014fundamental}
M.~Ji, G.~Caire, and A.~Molisch, ``Fundamental limits of caching in wireless
  d2d networks,'' \emph{IEEE Transactions on Information Theory}, vol.~62,
  no.~2, pp. 849 -- 869, 2014.

\bibitem{ji2016wireless}
M.~Ji, G.~Caire, and A.~F. Molisch, ``Wireless device-to-device caching
  networks: Basic principles and system performance,'' \emph{IEEE Journal on
  Selected Areas in Communications}, vol.~34, no.~1, pp. 176--189, 2016.

\bibitem{golrezaei2014base}
N.~Golrezaei, P.~Mansourifard, A.~F. Molisch, and A.~G. Dimakis, ``Base-station
  assisted device-to-device communications for high-throughput wireless video
  networks,'' \emph{IEEE Transactions on Wireless Communications}, vol.~13,
  no.~7, pp. 3665--3676, 2014.

\bibitem{PerabathiniBKDC15}
\BIBentryALTinterwordspacing
B.~Perabathini, E.~Bastug, M.~Kountouris, M.~Debbah, and A.~Conte, ``Caching at
  the {E}dge: a {G}reen {P}erspective for 5{G} {N}etworks,'' \emph{CoRR}, vol.
  abs/1503.05365, 2015. [Online]. Available:
  \url{http://arxiv.org/abs/1503.05365}
\BIBentrySTDinterwordspacing

\bibitem{sarzaidi2015}
S.~A.~R. Zaidi, M.~Ghogho, and D.~C. McLernon, ``Information {C}entric
  {M}odeling for {T}wo-tier {C}ache {E}nabled {C}ellular {N}etworks,''
  \emph{IEEE International Conference on Communications (ICC)}, 2015.

\bibitem{tamoor2015modeling}
S.~Tamoor-ul Hassan, M.~Bennis, P.~H. Nardelli, and M.~Latva-Aho, ``Modeling
  and {A}nalysis of {C}ontent {C}aching in {W}ireless {S}mall {C}ell
  {N}etworks,'' \emph{arXiv preprint arXiv:1507.00182}, 2015.

\bibitem{bastug2014cache}
E.~Bastug, M.~Bennis, and M.~Debbah, ``Cache-{E}nabled {S}mall {C}ell
  {N}etworks: {M}odeling and {T}radeoffs,'' in \emph{11th International
  Symposium on Wireless Communications Systems (ISWCS)}.\hskip 1em plus 0.5em
  minus 0.4em\relax IEEE, 2014, pp. 649--653.

\bibitem{afshang2015fundamentals}
M.~Afshang, H.~S. Dhillon, and P.~H.~J. Chong, ``Fundamentals of
  cluster-centric content placement in cache-enabled device-to-device
  networks,'' \emph{arXiv preprint arXiv:1509.04747}, 2015.

\bibitem{chen2016cooperative}
Z.~Chen, J.~Lee, T.~Q. Quek, and M.~Kountouris, ``Cooperative caching and
  transmission design in cluster-centric small cell networks,'' \emph{arXiv
  preprint arXiv:1601.00321}, 2016.

\bibitem{afshang2015modeling}
M.~Afshang, H.~S. Dhillon, and P.~H.~J. Chong, ``Modeling and {P}erformance
  {A}nalysis of {C}lustered {D}evice-to-{D}evice {N}etworks,'' \emph{arXiv
  preprint arXiv:1508.02668}, 2015.

\bibitem{AltieriPVG14}
\BIBentryALTinterwordspacing
A.~Altieri, P.~Piantanida, L.~R. Vega, and C.~G. Galarza, ``On {F}undamental
  {T}rade-offs of {D}evice-to-{D}evice {C}ommunications in {L}arge {W}ireless
  {N}etworks,'' \emph{CoRR}, vol. abs/1405.2295, 2014. [Online]. Available:
  \url{http://arxiv.org/abs/1405.2295}
\BIBentrySTDinterwordspacing

\bibitem{lin2014spectrum}
X.~Lin, J.~G. Andrews, and A.~Ghosh, ``Spectrum {S}haring for
  {D}evice-to-{D}evice {C}ommunication in {C}ellular {N}etworks,'' \emph{IEEE
  Transactions on Wireless Communications}, vol.~13, no.~12, pp. 6727--6740,
  2014.

\bibitem{elsawy2014analytical}
H.~ElSawy, E.~Hossain, and M.-S. Alouini, ``Analytical {M}odeling of {M}ode
  {S}election and {P}ower {C}ontrol for {U}nderlay d2d {C}ommunication in
  {C}ellular {N}etworks,'' \emph{IEEE Transactions on Communications}, vol.~62,
  no.~11, pp. 4147--4161, 2014.

\bibitem{niesen2014coded}
U.~Niesen and M.~A. Maddah-Ali, ``Coded {C}aching for {D}elay-{S}ensitive
  {C}ontent,'' \emph{arXiv preprint arXiv:1407.4489}, 2014.

\bibitem{dabirmoghaddam2014understanding}
A.~Dabirmoghaddam, M.~M. Barijough, and J.~Garcia-Luna-Aceves, ``Understanding
  {O}ptimal {C}aching and {O}pportunistic {C}aching at the {E}dge of
  {I}nformation-{C}entric {N}etworks,'' in \emph{Proceedings of the 1st
  international conference on Information-centric networking}.\hskip 1em plus
  0.5em minus 0.4em\relax ACM, 2014, pp. 47--56.

\bibitem{bacstuug2015transfer}
E.~Ba{\c{s}}tu{\u{g}}, M.~Bennis, and M.~Debbah, ``A {T}ransfer {L}earning
  {A}pproach for {C}ache-{E}nabled {W}ireless {N}etworks,'' \emph{arXiv
  preprint arXiv:1503.05448}, 2015.

\bibitem{fricker2012impact}
C.~Fricker, P.~Robert, J.~Roberts, and N.~Sbihi, ``Impact of {T}raffic {M}ix on
  {C}aching {P}erformance in a {C}ontent-{C}entric {N}etwork,'' in \emph{IEEE
  Conference on Computer Communications Workshops (INFOCOM WKSHPS)}.\hskip 1em
  plus 0.5em minus 0.4em\relax IEEE, 2012, pp. 310--315.

\bibitem{blaszczyszyn2015optimal}
B.~Blaszczyszyn and A.~Giovanidis, ``Optimal geographic caching in cellular
  networks,'' in \emph{IEEE International Conference on Communications
  (ICC)}.\hskip 1em plus 0.5em minus 0.4em\relax IEEE, 2015, pp. 3358--3363.

\bibitem{avrachenkov2016optimization}
K.~Avrachenkov, X.~Bai, and J.~Goseling, ``Optimization of caching devices with
  geometric constraints,'' \emph{arXiv preprint arXiv:1602.03635}, 2016.

\bibitem{LiuY15}
\BIBentryALTinterwordspacing
D.~Liu and C.~Yang, ``Energy {E}fficiency of {D}ownlink {N}etworks with
  {C}aching at {B}ase {S}tations,'' \emph{CoRR}, vol. abs/1505.06615, 2015.
  [Online]. Available: \url{http://arxiv.org/abs/1505.06615}
\BIBentrySTDinterwordspacing

\bibitem{stoyan1987stochastic}
D.~Stoyan, W.~S. Kendall, J.~Mecke, and L.~Ruschendorf, \emph{Stochastic
  geometry and its applications}.\hskip 1em plus 0.5em minus 0.4em\relax Wiley
  New York, 1987, vol.~2.

\bibitem{yu2013downlink}
S.~M. Yu and S.-L. Kim, ``Downlink capacity and base station density in
  cellular networks,'' in \emph{11th International Symposium on Modeling \&
  Optimization in Mobile, Ad Hoc \& Wireless Networks (WiOpt)}.\hskip 1em plus
  0.5em minus 0.4em\relax IEEE, 2013, pp. 119--124.

\bibitem{andrews2011tractable}
J.~G. Andrews, F.~Baccelli, and R.~K. Ganti, ``A tractable approach to coverage
  and rate in cellular networks,'' \emph{IEEE Transactions on Communications},
  vol.~59, no.~11, pp. 3122--3134, 2011.

\bibitem{foss1996voronoi}
S.~Foss and S.~Zuyev, ``On a {V}oronoi {A}ggregative {P}rocess {R}elated to a
  {B}ivariate {P}oisson {P}rocess,'' \emph{Advances in Applied Probability},
  pp. 965--981, 1996.

\bibitem{aafzal2016}
A.~Afzal, S.~A.~R. Zaidi, M.~Ghogho, and D.~C. McLernon, ``On the analysis of
  cellular networks with caching and coordinated device-to-device
  communication,'' \emph{IEEE International Conference on Communications
  (ICC)}, 2016.

\bibitem{moltchanov2012survey}
D.~Moltchanov, ``Survey paper: Distance distributions in random networks,''
  \emph{Ad Hoc Networks}, vol.~10, no.~6, pp. 1146--1166, 2012.

\bibitem{weisstein2003circle}
E.~W. Weisstein, ``Circle-circle intersection,'' 2003.

\end{thebibliography}

\end{document}